\documentclass[aps,twocolumn,prd,showpacs,showkeys,preprintnumbers,superscriptaddress,bibnotes,floatfix,longbibliography,notitlepage,nofootinbib]{revtex4-2}

\pdfoutput=1
\usepackage{amsmath}
\usepackage{amsfonts}
\usepackage{amssymb}
\usepackage{mathrsfs}
\usepackage{graphicx}
\usepackage{color}
\usepackage{bbding}
\usepackage{tabularx}
\usepackage[usenames,dvipsnames,table]{xcolor}
\usepackage[normalem]{ulem}
\usepackage{makecell}
\usepackage{siunitx}
\usepackage{enumerate}
\usepackage{multirow}
\usepackage{makecell}
\usepackage{upgreek}
\definecolor{darkred}{rgb}{0.5,0,0}
\definecolor{darkblue}{rgb}{0,0,0.5}
\definecolor{firebrick}{rgb}{0.75,0.125,0.125}
\definecolor{darkgreen}{rgb}{0,0.5,0}
\usepackage[colorlinks=true,linkcolor=firebrick,citecolor=darkgreen,urlcolor=darkblue]{hyperref}
\usepackage{fontawesome5}

\newcommand{\orcid}[1]{\href{https://orcid.org/#1}{\includegraphics[width=10pt]{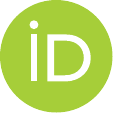}}}

\newcommand{\ie}{{\it i.e.}}

\newcommand{\eg}{{\it e.g.}}

\newcommand{\cf}{{\it cf.}}

\newcommand{\eq}{Eq.}

\newcommand{\fig}{Fig.}

\newcommand{\Refe}{Ref.}
\newcommand{\Refes}{Refs.}
\newcommand{\equ}[1]{\eq~(\ref{equ:#1})}
\newcommand{\figu}[1]{\fig~\ref{fig:#1}}
\newcommand{\exclude}[1]{{}}
\long\def\exclude#1{}

\begin{document}

\title{New limits on neutrino decay from high-energy astrophysical neutrinos~\href{https://github.com/damianofiorillo/Neutrino\_Decay\_Data}{\faGithubSquare}}

\author{V\'ictor B.~Valera
\orcid{0000-0002-0532-5766}}
\email{vvalera@nbi.ku.dk}
\affiliation{Niels Bohr International Academy, Niels Bohr Institute,\\University of Copenhagen, DK-2100 Copenhagen, Denmark}

\author{Damiano F.~G.~Fiorillo
\orcid{0000-0003-4927-9850}}
\email{damiano.fiorillo@nbi.ku.dk}
\affiliation{Niels Bohr International Academy, Niels Bohr Institute,\\University of Copenhagen, DK-2100 Copenhagen, Denmark}

\author{Ivan Esteban
\orcid{0000-0001-5265-2404}}
\email{ivan.esteban@ehu.eus}
\affiliation{Department of Physics, University of the Basque Country UPV/EHU, PO Box 644, 48080 Bilbao, Spain}
\affiliation{EHU Quantum Center, University of the Basque Country UPV/EHU}

\author{Mauricio Bustamante
\orcid{0000-0001-6923-0865}}
\email{mbustamante@nbi.ku.dk}
\affiliation{Niels Bohr International Academy, Niels Bohr Institute,\\University of Copenhagen, DK-2100 Copenhagen, Denmark}

\date{\today}

\begin{abstract}
 Since neutrinos have mass differences, they could decay into one another.  But their lifetimes are likely long, even when shortened by new physics, so decay likely impacts neutrinos only during long trips.  This makes high-energy astrophysical neutrinos, traveling for up to billions of light-years, sensitive probes of decay.  However, their sensitivity must be tempered by reality.  We derive from them thorough bounds on the neutrino lifetimes accounting for critical astrophysical unknowns and the nuances of neutrino detection.  Using the diffuse neutrino flux, we disfavor lifetimes $\tau \lesssim 20$--450~s~$(m/{\rm eV})$, based on present IceCube data, and forecast factor-of-10 improvements by upcoming detectors.  Using, for the first time, neutrinos from the active galaxy NGC 1068, extant unknowns preclude placing lifetime bounds today, but upcoming detectors could disfavor $\tau \sim 100$--5000~s~$(m/{\rm eV})$.
\end{abstract}

\maketitle


\section{Introduction}\label{sec:introduction}

The discovery of neutrino oscillations showed that different neutrino states have different masses.  This, in turn, implies that they might decay into each other. In minimal extensions of the Standard Model that accommodate massive neutrinos, their lifetime is capped by radiative decays, but is predicted to be vastly longer than the age of the Universe~\cite{Pal:1981rm, Hosotani:1981mq, Nieves:1982bq}, making them effectively stable.  However, in non-minimal extensions, neutrinos may decay much faster via new channels~\cite{Bahcall:1972my, Shrock:1974nd, Petcov:1976ff, Marciano:1977wx, Zatsepin:1978iy, Chikashige:1980qk, Gelmini:1980re, Pal:1981rm, Schechter:1981cv, Shrock:1982sc, Gelmini:1983ea, Bahcall:1986gq, Nussinov:1987pc, Frieman:1987as, Kim:1990km} and lead to effects that may be detectable within shorter time scales.  Thus, looking for neutrino decay may guide us in extending the Standard Model.  

A neutrino $\nu_i$ might decay into a lighter one, $\nu_j$, plus an additional particle, $\Phi$, \ie, $\nu_i \to \nu_j + \Phi$; we elaborate on this in Sec.~\ref{sec:nu_prop_with_decay}.  As a result, in a flux of neutrinos of energy $E_\nu$, mass $m_i$, and lifetime $\tau_i$, the effect of decay a time $t$ after neutrino emission is given by $e^{-\frac{t}{\gamma \tau}} = e^{-\frac{m_i}{\tau_i} \frac{L}{E_\nu}}$, where, because we deal with relativistic neutrinos, $L \approx t$ is their traveled distance and $\gamma \equiv E_\nu / m_i$, their Lorentz boost.  This represents a depletion of the flux due to the disappearance of decaying neutrinos [and its enhancement due to the appearance of daughter neutrinos, which we do not consider (Sec.~\ref{sec:nu_prop_with_decay})].  Because presently we do not know the values of the neutrino masses, we are sensitive only to the combination $\tau_i / m_i$, which we refer to below as the ``lifetime'' (see, however, \Refe~\cite{Pompa:2023yzg}).  

\begin{figure}[t!]
 \centering
 \includegraphics[width=0.48\textwidth]{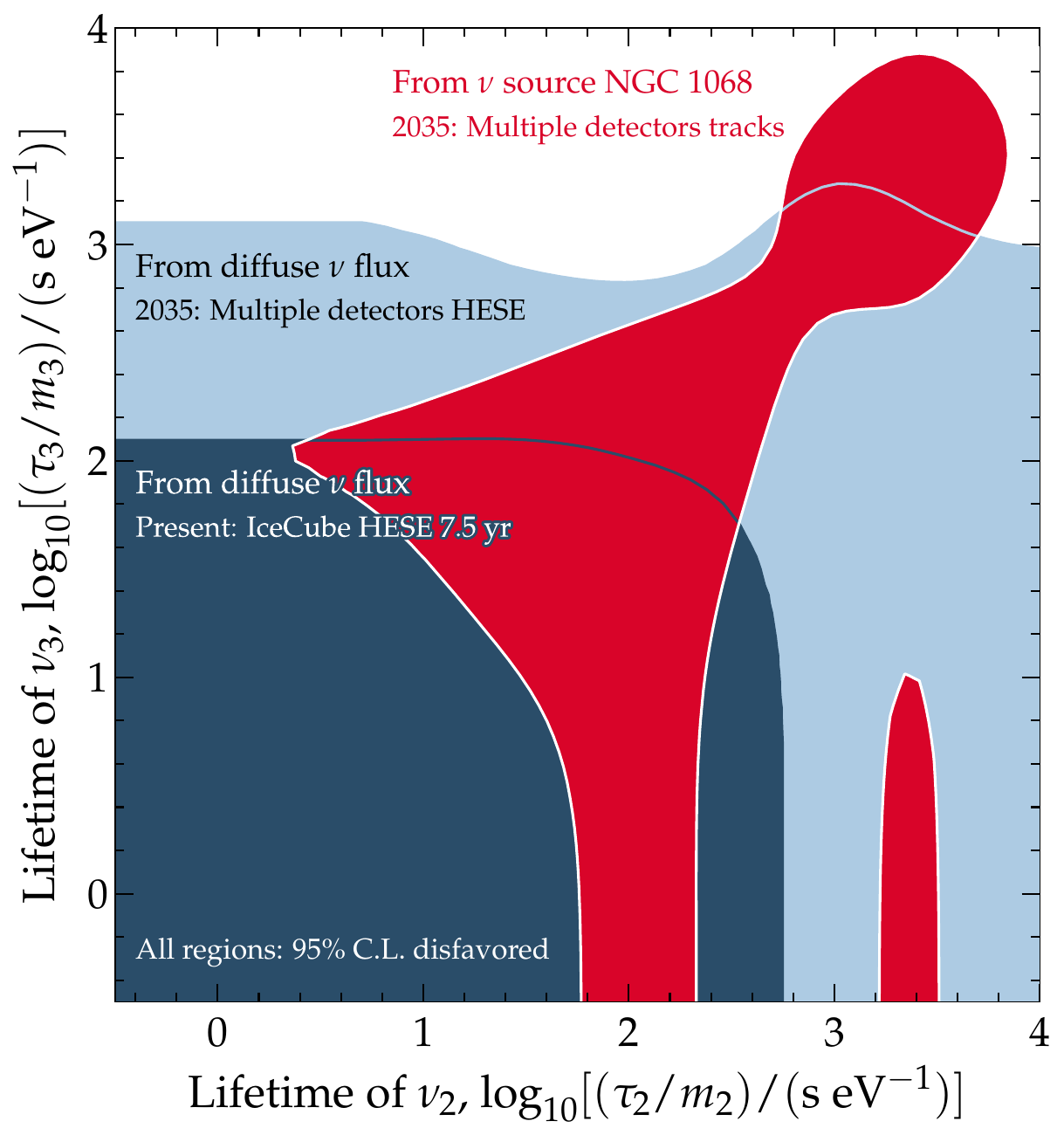}
 \caption{\textbf{\textit{Overview of bounds on neutrino lifetimes from TeV--PeV astrophysical neutrinos.}}  Decay is of the mass eigenstates $\nu_2$ and $\nu_3$ into invisible products, which modifies the neutrino energy spectrum and flavor composition.
 Present bounds are from the diffuse neutrino flux, using the IceCube 7.5-year High-Energy Starting Events (HESE) sample; no bound can be placed using present observations of the neutrino source NGC 1068.  Forecasts are for 2035 using the combined detection by IceCube and upcoming detectors Baikal-GVD, IceCube-Gen2, KM3NeT, P-ONE, TAMBO, and TRIDENT.  In this figure, the bounds are obtained using baseline astrophysical assumptions;
 Figs.~\ref{fig:bounds_ngc1068} and \ref{fig:bounds_diffuse} show alternatives. 
 \textit{Bounds derived from the diffuse flux are robust to uncertainty in astrophysical modeling (\figu{bounds_diffuse}), but bounds derived from a point neutrino source are strongly impacted by it (\figu{bounds_ngc1068}).}}
 \vspace*{-0.7cm}
 \label{fig:bounds_overview}
\end{figure}

Upon detection, a significant fraction of $\nu_i$ will have decayed if $\tau_i / m_i \lesssim L / E_\nu$.  Although neutrino decay remains unobserved, neutrino lifetimes have been constrained by searches using neutrinos from natural and human-made sources~\cite{Super-Kamiokande:2006jvq, MINOS:2006foh, K2K:2006yov, Gonzalez-Garcia:2008mgl, Gomes:2014yua, Abrahao:2015rba, Choubey:2017dyu, Choubey:2018cfz, deSalas:2018kri, Abdullahi:2020rge, Ackermann:2022rqc, Ternes:2024qui, Martinez-Mirave:2024hfd} that look for signs of decay in the neutrino energy spectrum and flavor composition, \ie, the proportion of $\nu_e$, $\nu_\mu$, and $\nu_\tau$ in an observed flux of neutrinos.  In neutrino experiments with terrestrial-scale baselines, of up to a few thousand km, significant time dilation of the neutrino lifetime in the laboratory frame limits our sensitivity to $\lesssim 10^{-10}$~s~eV~$^{-1}$~\cite{Gonzalez-Garcia:2008mgl, Abrahao:2015rba}. 

Astrophysical neutrinos overcome this limitation by virtue of their longer baselines.  Solar neutrinos, with MeV-scale energies and a baseline of 1~A.U., probe lifetimes of around $5 \cdot 10^{-4}$~s~eV$^{-1}$~\cite{Beacom:2002cb, Berryman:2014qha}.  Neutrinos from supernova SN~1987A~\cite{Kamiokande-II:1987idp, Bionta:1987qt, Alekseev:1988gp}, with tens of MeV and a baseline of $50$~kpc, probe lifetimes of around $5 \cdot 10^5$~s~eV$^{-1}$ (though they are subject to  model uncertainties)~\cite{Ivanez-Ballesteros:2023lqa, Martinez-Mirave:2024hfd}.  

The IceCube discovery~\cite{IceCube:2013low} of neutrinos with TeV--PeV energies and Mpc--Gpc baselines has allowed us to probe lifetimes of $\tau_i / m_i \sim 10^3 (L/{\rm Gpc}) (100~{\rm TeV}/E_\nu)$~s~eV$^{-1}$, matching or outperforming the above searches.  
Yet, so far, as pointed out by \Refe~\cite{Bustamante:2016ciw}, searches for neutrino decay with high-energy astrophysical neutrinos~\cite{Shoemaker:2015qul, Bustamante:2016ciw, Denton:2018aml} have been hampered, implicitly or explicitly, by severely incomplete information on the neutrino properties, source properties, and detection aspects.  The presence of these uncertainties, often overlooked, have made it difficult to assess how robust are the bounds on the neutrino lifetime derived from high-energy astrophysical neutrinos.

As high-energy neutrino physics matures, the above limitations weaken.  But, more importantly, our ability to quantify their impact on searches for neutrino decay---and searches for other new physics---grows.  Recent years have seen improvement in our knowledge of neutrino properties---masses and mixing parameters---a growing number of detected high-energy neutrinos, and improvement in reconstruction techniques.  Today, significant uncertainties still affect the searches for decay, but, compared to earlier studies, we are better equipped to assess their impact thanks to the above developments and to the public availability of experimental neutrino data and detailed detector simulations.

Motivated by this, we revisit neutrino decay with high-energy astrophysical neutrinos from a fresh perspective.  We provide new limits on neutrino lifetimes; arguably, these are the most robust limits based on high-energy astrophysical  neutrinos to date.  But our goal is broader: to show the manifest impact of present-day unknowns, the realistic opportunities made possible by spectacular recent discoveries---notably, of the first steady-state high-energy astrophysical neutrino source, NGC 1068~\cite{IceCube:2022der}---and the near-future reach of new detectors.

The rest of this paper is organized as follows.  In Sec.~\ref{sec:synopsis} we outline the new perspectives of our work.  In Sec.~\ref{sec:nu_prop_with_decay} we discuss the phenomenological impact of neutrino decay on the observed neutrino spectrum and flavor composition. In Sec.~\ref{sec:bounds_ngc1068}, we discuss the bounds that can be set using the neutrino signal from NGC 1068, and in Sec.~\ref{sec:bounds_diffuse} we obtain bounds from the diffuse neutrino flux. Finally, in Sec.~\ref{sec:discussion} we summarize and discuss our results.


\section{Synopsis}
\label{sec:synopsis}

In contrast to previous searches for neutrino decay that use high-energy astrophysical neutrinos, ours embraces hitherto unexplored or partially explored uncertainties and model neutrino detection in a realistic way:
\begin{description}
 \item[Astrophysical uncertainties] Because the bulk of the sources and the production mechanisms of high-energy astrophysical neutrino are unknown, large uncertainties in the predictions of the neutrino flux cloud the presence of potential features introduced by neutrino decay. We explore the impact on the sensitivity to decay of unknowns in the size of the emitted neutrino flux, its energy spectrum, flavor composition, the distribution of neutrino sources, and the number of source populations.
 \item[Detailed neutrino detection] The precision with which high-energy neutrino telescopes infer the energy, direction, and flavor of detected neutrinos limits our capability to spot potential features introduced by neutrino decay.  To account for this, we model high-energy neutrino detection in as much detail as is publicly accessible, including energy and direction resolution, in-Earth propagation effects, and backgrounds of atmospheric neutrinos and muons.  We use Monte Carlo event samples provided by the IceCube Collaboration when available~\cite{IC75yrHESEPublicDataRelease} (\ie, for the diffuse flux below), and third-party approximations otherwise~\cite{PlenumCode} (\ie, for the point-source flux below).
\end{description}
The impact of the above perspectives on the sensitivity to decay ranges from appreciable to critical.  They render our results realistic, but also weaken them; in some cases---in particular, for neutrinos from an identified source (see below)---to the point of nearly eliminating sensitivity to neutrino decay altogether.  We elaborate on the above perspectives later.

We apply these perspectives along two fronts:
\begin{description}
 \item[Diffuse neutrino flux]
  We revisit tests of neutrino decay that use the diffuse flux of high-energy astrophysical neutrinos, \ie, the aggregated flux from all unresolved sources.  So far, they had been only estimated in \Refe~\cite{Bustamante:2016ciw} (see also \Refes~\cite{Pakvasa:1981ci, Beacom:2002vi, Barenboim:2003jm, Beacom:2003nh, Beacom:2003zg, Meloni:2006gv, Maltoni:2008jr, Bustamante:2010nq, Baerwald:2012kc, Pakvasa:2012db, Dorame:2013lka, Pagliaroli:2015rca, Bustamante:2015waa, Huang:2015flc, Shoemaker:2015qul}).  \textbf{\textit{For the first time, we report bounds on the neutrino lifetime from the diffuse neutrino flux rigorously derived from a nuanced detector description.}}  We look for signs of decay jointly in the energy spectrum and the flavor composition of the neutrino flux, as proposed by \Refe~\cite{Shoemaker:2015qul}.
 \item[Point-source neutrino flux]
  Recently, IceCube discovered the first steady-state astrophysical candidate source of high-energy neutrinos, the active galaxy NGC 1068~\cite{IceCube:2022der}, located 14~Mpc away.  Unlike previously reported transient candidate sources~\cite{IceCube:2018cha, IceCube:2018dnn, Stein:2020xhk, Reusch:2021ztx}, the number of neutrinos detected from NGC 1068 is a few tens, enough to make it plausible to search for the energy-dependent features characteristic of neutrino decay (Sec.~\ref{sec:nu_prop_with_decay}).  Given the angular resolution with which IceCube detects neutrinos from NGC 1068---and from any other possible extragalactic neutrino source---it is effectively a point source. \textbf{\textit{For the first time, we search for signs of decay in the neutrinos from NGC 1068.}}  We do so using only their energy spectrum, since their flavor composition has not yet been measured (see, however, \Refe~\cite{Valera:2023bud}).
\end{description}
We report bounds on neutrino lifetime based on present-day observations by IceCube and on projected observations by its combination with upcoming or planned neutrino telescopes Baikal-GVD~\cite{Baikal-GVD:2023beh}, IceCube-Gen2~\cite{IceCube-Gen2:2020qha}---the envisioned high-energy upgrade of IceCube---KM3NeT~\cite{KM3Net:2016zxf}, P-ONE~\cite{P-ONE:2020ljt}, TAMBO~\cite{Thompson:2023pnl} , and TRIDENT~\cite{Ye:2022vbk}.  We expand on them later.

Figure~\ref{fig:bounds_overview} shows an overview of our present-day and projected bounds obtained under baseline model choices.  We defer details about them to Secs.~\ref{sec:bounds_ngc1068} and \ref{sec:bounds_diffuse}.   We consider the non-radiative decay of the mass eigenstates $\nu_2$ and $\nu_3$ into $\nu_1$, which we assume to be the lightest one, plus an undetectable partner.  As we discuss below (Sec.~\ref{sec:nu_prop_with_decay-overview_decay}), we focus as a benchmark on invisible decay, where the daughter $\nu_1$ is also undetectable.  

Presently, the diffuse neutrino flux constrains the lifetimes to be smaller than a few times 100~s~eV$^{-1}$.  Future combined detection by multiple detectors in the year 2035 could improve this by one order of magnitude for $\nu_3$, assuming that the real flavor composition is the one suggested by present-day data (more on this later).  The projected bounds from diffuse flux are robust: \textbf{\textit{they are weakened by adopting alternative model choices within the breadth of astrophysical uncertainties, but do not vanish.}}

In contrast, present observations of NGC 1068 cannot constrain the neutrino lifetime, but \figu{bounds_overview} shows that future observations may probe lifetimes of $10^3$--$10^4$~s~eV$^{-1}$, beyond the reach of the bounds from the diffuse flux.  However, we find that this is true only under favorable model choices, like those made in \figu{bounds_overview}.  \textbf{\textit{Under most other model choices within the breadth of astrophysical uncertainties the projected lifetime bounds from NGC 1068 vanish or nearly vanish.}} 

The above general observations, which we expand upon below, illustrate a larger point: for searches for physics beyond the Standard Model that use high-energy astrophysical neutrinos to be robust, they must be tempered by realistic astrophysical uncertainties.


\section{Decay in high-energy astrophysical neutrinos}
\label{sec:nu_prop_with_decay}


\subsection{Overview of invisible neutrino decay}
\label{sec:nu_prop_with_decay-overview_decay}

If neutrinos are unstable, high-energy neutrinos emitted by an astrophysical source may decay on their way to Earth.  If the decay is significant, \ie, if the neutrino lifetimes are sufficiently short, it may modify the energy spectrum and flavor composition of the neutrino flux that reaches Earth.  Below, we describe these modifications; later (Secs.~\ref{sec:bounds_ngc1068} and \ref{sec:bounds_diffuse}), we look for evidence of them.  

The magnitude of the modifications introduced by decay depends on the size of the neutrino lifetime, $\tau_i/m_i$, relative to $L/E_\nu$ (Sec.~\ref{sec:introduction}), but their specific form depends on how neutrinos decay.  We focus on the case of ``invisible decay'', where neutrinos decay to particles that are undetected by high-energy neutrino telescopes, \eg, low-energy neutrinos or non-standard particles like Majorons and sterile neutrinos.  This is not the most general decay scenario, but it is a useful benchmark, since it captures the essential features of decay and allows us to clearly show the impact of the new perspectives introduced in Sec.~\ref{sec:synopsis}.  We comment on alternative scenarios later.

We consider three active neutrino flavor states, $\nu_e$, $\nu_\mu$, and $\nu_\tau$, each a superposition of the neutrino mass eigenstates $\nu_1$, $\nu_2$, and $\nu_3$, \ie, $\nu_\alpha = \sum_{i=1}^3 U_{\alpha i}^\ast \nu_i$, ($\alpha = e, \mu, \tau$ and $i = 1, 2, 3$), where $U_{\alpha i}$ are elements of the Pontecorvo-Maki-Nakagawa-Sakata (PMNS) mixing matrix~\cite{Pontecorvo:1957qd,Maki:1962mu}.  It is the mass eigenstates that decay, since, unlike flavor states, they have a well-defined mass.  We explore the invisible decay of $\nu_2$ and $\nu_3$, while keeping $\nu_1$ stable.  This is motivated by the present mild preference for the normal neutrino mass ordering, where $\nu_1$ is the lightest of the mass eigenstates, but there are other possible decay channels~\cite{Mehta:2011qb, Abdullahi:2020rge}.  (The scenario in which $\nu_1$ is unstable is  constrained by the observation of neutrinos from supernova SN 1987A~\cite{Martinez-Mirave:2024hfd}.  An unstable $\nu_1$ demands a total energy emitted in neutrinos larger than seen in even the most optimistic supernova simulations~\cite{Fiorillo:2023frv}.)

In the more general case of ``visible decay'', which we do not explore, the heavier neutrino mass eigenstates decay into the lightest one, \eg, $\nu_2, \nu_3 \to \nu_1 + \Phi$, where $\nu_1$ is detectable and $\Phi$ is, in general, not.  The effect of decay on the neutrino energy spectrum depends on the fraction of the parent neutrino energy received by the daughter neutrino, and the flavor composition at Earth depends on the mass differences between parent and daughter neutrinos; see, \eg, \Refes~\cite{Bustamante:2016ciw, Abdullahi:2020rge}.  We also do not consider the possibility of chains of decay, \eg, $\nu_3 \to \nu_2 + \Phi$ followed by $\nu_2 \to \nu_1 + \Phi$; see \Refe~\cite{Abdullahi:2020rge} for a full treatment.  The conclusions that we draw later under invisible decay apply to neutrino decay generally, though the values of the limits on neutrino lifetime that we obtain do not.

The effects introduced by the decay of $\nu_2$ and $\nu_3$ depend on the factors $e^{-\frac{\tau_2}{m_2}\frac{L}{E_\nu}}$ and $e^{-\frac{\tau_3}{m_3}\frac{L}{E_\nu}}$ (Sec.~\ref{sec:introduction}).  Below, we show how they modify the flavor composition (Sec.~\ref{sec:nu_prop_with_decay-flavor_comp}) and energy spectrum (Sec.~\ref{sec:nu_prop_with_decay-diffuse_vs_point}) of high-energy astrophysical neutrinos.


\subsection{Neutrino decay in the flavor composition}
\label{sec:nu_prop_with_decay-flavor_comp}

The mechanism by which high-energy neutrinos are produced determines the flavor composition with which they are emitted, \ie, the proportion of $\nu_e$, $\nu_\mu$, and $\nu_\tau$ in their emitted flux.  Because neutrinos oscillate on their way to Earth, the flavor composition of the flux that reaches Earth is different from the one at production.  And if, in addition, neutrinos decay along the way, their flavor composition at Earth might deviate from the standard-oscillation expectation~\cite{Bustamante:2015waa}.

\smallskip

\textbf{\textit{Neutrino production.---}}Sources of high-energy astrophysical neutrinos are believed to produce them in interactions of high-energy protons with ambient matter or radiation (Sec.~\ref{sec:bounds_ngc1068}).  Both types of interaction produce high-energy charged pions---and other products---that, upon decaying, produce high-energy neutrinos.  

Pions decay via $\pi^+\to \mu^+ \nu_\mu$ followed by $\mu^+\to \overline{\nu}_\mu \nu_e e^+$, and their charge-conjugated processes.  Thus, sources (S) where this decay chain develops in full emit neutrinos with a \textit{pion-beam} flavor composition of $(f_e, f_\mu, f_\tau)_{\rm S} =(\frac{1}{3}, \frac{2}{3}, 0)$, where $f_{\alpha, {\rm S}}$ is the fraction of $\nu_\alpha + \bar{\nu}_\alpha$ in the flux.  In sources that harbor strong magnetic fields, the intermediate muons may cool via synchrotron radiation, so that, their energies dampened, the neutrinos they decay into are high-energy no more.  This yields a \textit{muon-damped} flavor composition of $(0, 1, 0)_{\rm S}$.  Finally, towards low energies, $\bar{\nu}_e$ from the beta-decay of neutrons co-produced with the pions could yield a \textit{neutron-beam} flavor composition of $(1, 0, 0)_{\rm S}$.  

We adopt the pion-beam, muon-damped, and neutron-beam scenarios as benchmarks in our analysis.  They---and all standard neutrino production scenarios---share $f_{\tau, {\rm S}} = 0$, since $\nu_\tau$ production has a large threshold center-of-mass energy to produce the taus from whose decays $\nu_\tau$ would be emitted.  There are other possibilities for the flavor composition emitted by the sources, including their evolving with neutrino energy, that depend, \eg, on the geometry and density of the sources and the energies reached by protons and other nuclei inside them; see, \eg, \Refes~\cite{Mucke:1999yb, Kashti:2005qa, Lipari:2007su, Hummer:2010ai, Hummer:2010vx, Baerwald:2010fk, Baerwald:2011ee, Winter:2012xq, Roulet:2020yye, Fiorillo:2021hty}.  In our analysis, rather than exploring them individually, we consider generic scenarios of flavor composition that capture the uncertainty in $f_{\alpha, {\rm S}}$.

\smallskip

\textbf{\textit{Flavor composition at Earth.---}}As a result of neutrino oscillations, the flavor composition at Earth ($\oplus$) is, for a given choice of $f_{\alpha, {\rm S}}$, 
\begin{equation}
 f_{\alpha, \oplus}
 =
 \sum_{\beta=e,\mu,\tau}
 \left(
 \sum_{i=1}^3
 \left\vert U_{\beta i} \right\vert^2
 \left\vert U_{\alpha i} \right\vert^2
 \right)
 f_{\beta, {\rm S}} \;,
\end{equation}
where the term in parentheses is the average probability of the flavor transition $\nu_\beta \to \nu_\alpha$, computed for high-energy astrophysical neutrinos (see, \eg, \Refes~\cite{Farzan:2008eg, Winter:2012xq, Ackermann:2022rqc, Fiorillo:2024jqz}).  The PMNS matrix elements depend on three mixing angles, $\theta_{12}$, $\theta_{23}$, and $\theta_{13}$, and one CP-violation phase, $\delta_{\rm CP}$, whose values are known from oscillation experiments to different degrees of precision.  In our analysis, we use their best-fit values and allowed ranges from the recent NuFIT~5.2 global fit to oscillation data~\cite{Esteban:2020cvm, NuFit5.2}, obtained assuming normal neutrino mass ordering.  (Using instead the values obtained under inverted ordering would not change our results significantly; see, \eg, \Refe~\cite{Bustamante:2015waa}.) 

\begin{figure}
 \centering
 \includegraphics[width=\columnwidth]{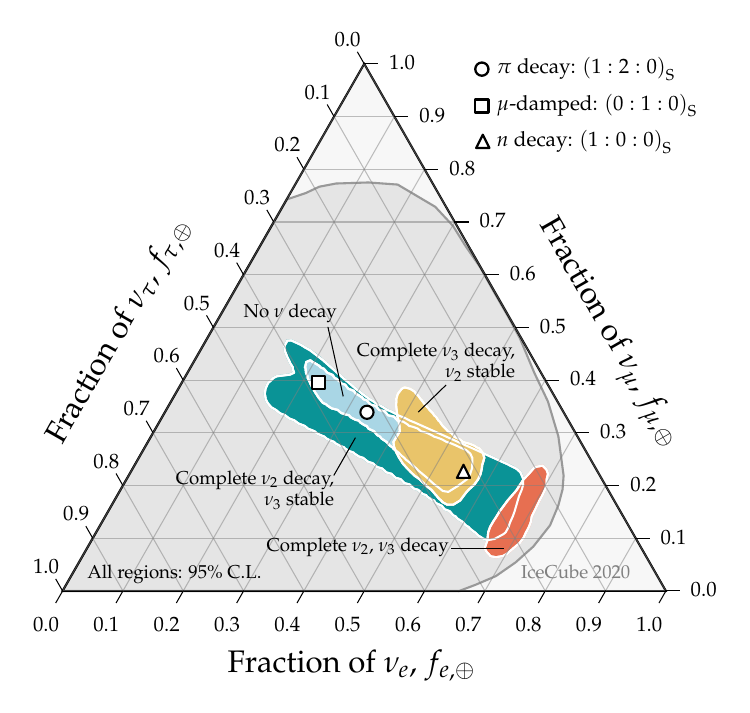}
 \caption{\textbf{\textit{Flavor composition of high-energy astrophysical neutrinos at Earth under the effects of neutrino decay.}} The regions of allowed flavor composition are generated by exploring all possible flavor combinations at the sources assuming no $\nu_\tau$ production, \ie, $(f_{e, {\rm S}}, f_{\mu, {\rm S}} = 1-f_{e, {\rm S}}, f_{\tau, {\rm S}} = 0)$, and accounting for present-day uncertainty on the neutrino mixing parameters, from NuFIT~5.2~\cite{Esteban:2020cvm, NuFit5.2}; see also \Refes~\cite{Bustamante:2015waa, Shoemaker:2015qul, Bustamante:2016ciw, Rasmussen:2017ert, Song:2020nfh, Liu:2023flr}.  Separate regions are for the standard case of no neutrino decay, invisible decay of $\nu_2$ only, of $\nu_3$ only, and of both, in all cases into $\nu_1$ and an undetected partner.  For comparison, we show expectations for three benchmark production scenarios---pion decay, muon-damped pion decay, and neutron decay---and the latest flavor measurement by IceCube~\cite{IceCube:2020fpi}.  \textit{Because the region of flavor composition under complete decay of $\nu_2$, or of $\nu_2$ and $\nu_3$, is clearly separated from the standard region and near the edge of the IceCube measured region, our analysis (Sec.~\ref{sec:bounds_diffuse}) using the present-day diffuse neutrino flux already disfavors it.}}
 \label{fig:flavor_triangle}
\end{figure}

Figure~\ref{fig:flavor_triangle} shows the flavor composition at Earth corresponding to our three benchmark scenarios, computed assuming the present-day best-fit values of the mixing parameters.  The nominal expectation, from the pion-beam scenario, yields approximately equal proportion of each flavor at Earth, \ie, roughly $\left( \frac{1}{3}, \frac{1}{3}, \frac{1}{3} \right)_\oplus$.  

Figure~\ref{fig:flavor_triangle} also shows the \textit{theoretically palatable region} of flavor composition at Earth~\cite{Bustamante:2015waa} (see also \Refes~\cite{Barenboim:2003jm, Arguelles:2015dca, Song:2020nfh, Liu:2023flr}).  This is the region of flavor composition generated by simultaneously varying the values of the mixing parameters within their allowed ranges and the flavor composition at the sources as $(f_{e, {\rm S}}, f_{\mu, {\rm S}} \equiv 1-f_{e,{\rm S}}, f_{\tau,{\rm S}} = 0)$, with $0 \leq f_{e, {\rm S}} \leq 1$.  As stated in \Refe~\cite{Bustamante:2015waa}, the length of the long axis of the region is due mainly to the variation in $f_{e, {\rm S}}$, and also to the uncertainty in $\theta_{23}$ and $\delta_{\rm CP}$; the length of the short axis is due to the uncertainty in $\theta_{12}$.  The effect of the uncertainty in $\theta_{13}$ is tiny.  (Allowing $f_{\tau, {\rm S}} \neq 0$ enlarges the region only marginally~\cite{Bustamante:2015waa}.)

Today, the uncertainties in the values of the mixing parameters are small enough to render the theoretically palatable region relatively small, about 10\% of the flavor triangle in \figu{flavor_triangle}.  However, they are still sizable enough to confound effects of neutrino decay, as we show next. 

The flavor composition is a versatile probe of neutrino astrophysics and fundamental physics.  For astrophysics, it reflects the conditions present in the astrophysical neutrino sources and so may help narrow down their identity~\cite{Rachen:1998fd, Athar:2000yw, Crocker:2001zs, Barenboim:2003jm, Beacom:2003nh, Beacom:2004jb, Kashti:2005qa, Mena:2006eq, Kachelriess:2006ksy, Lipari:2007su, Esmaili:2009dz, Choubey:2009jq, Hummer:2010ai, Bustamante:2010bf, Winter:2013cla, Palladino:2015zua, Bustamante:2015waa, Biehl:2016psj, Bustamante:2019sdb, Ackermann:2019ows, Bustamante:2020bxp, Song:2020nfh, Bhattacharya:2023mmp, Telalovic:2023tcb, Dev:2023znd, Capanema:2024hdm}.  For fundamental physics, it probes neutrino mixing and may reveal new physics~\cite{Beacom:2002vi, Barenboim:2003jm, Beacom:2003nh, Beacom:2003eu, Beacom:2003zg, Serpico:2005bs, Mena:2006eq, Lipari:2007su, Pakvasa:2007dc, Esmaili:2009dz, Choubey:2009jq, Esmaili:2009fk, Bhattacharya:2009tx, Bhattacharya:2010xj, Bustamante:2010nq, Mehta:2011qb, Baerwald:2012kc, Fu:2012zr, Pakvasa:2012db, Chatterjee:2013tza, Xu:2014via, Aeikens:2014yga, Arguelles:2015dca, Bustamante:2015waa, Pagliaroli:2015rca, Shoemaker:2015qul, deSalas:2016svi, Gonzalez-Garcia:2016gpq, Bustamante:2016ciw, Rasmussen:2017ert, Dey:2017ede, Bustamante:2018mzu, Farzan:2018pnk, Ahlers:2018yom, Brdar:2018tce, Palladino:2019pid, Ackermann:2019cxh, Arguelles:2019rbn, Ahlers:2020miq, Karmakar:2020yzn, Fiorillo:2020gsb, Song:2020nfh, Arguelles:2022tki, MammenAbraham:2022xoc, Telalovic:2023tcb}. 
Today, the IceCube neutrino telescope regularly measures the flavor composition of high-energy astrophysical neutrinos~\cite{Mena:2014sja, Palomares-Ruiz:2015mka, IceCube:2015rro, Palladino:2015vna, IceCube:2015gsk, Vincent:2016nut, IceCube:2020fpi}; we elaborate on this later (Sec.~\ref{sec:bounds_diffuse-hese}).

\smallskip

\textbf{\textit{Decay in the flavor composition.---}}The different mass eigenstates, $\nu_1$, $\nu_2$, and $\nu_3$, have different flavor content.  The content of $\nu_\alpha$ in $\nu_i$ is $\left\vert U_{\alpha i} \right\vert^2$, with the uncertainty on its value due to the uncertainty in the mixing parameters.  Roughly, $\nu_1$ has 68\% of electron flavor in it, $\nu_2$ has 30\%, and $\nu_3$ has 3\%, and the remaining flavor content in each is divided equally between muon and tau; see Fig.~1 in \Refe~\cite{Bustamante:2015waa}, Fig.~2 in \Refe~\cite{Bustamante:2016ciw}, and Fig.~5 in \Refe~\cite{Song:2020nfh}.
Thus, because the invisible decay of $\nu_2$ and $\nu_3$ changes their abundance relative to $\nu_1$, it alters the flavor composition of the high-energy neutrino flux at Earth.  

In the extreme case of complete decay of $\nu_2$ and $\nu_3$ upon reaching Earth, only $\nu_1$ remain, and so the flavor composition of the flux is given solely by the flavor content of $\nu_1$, \ie, $f_{\alpha, \oplus} = \left\vert U_{\alpha 1} \right\vert$, regardless of the flavor composition at the sources.  When the decay is incomplete, the flavor composition at the Earth depends on the fraction of remaining $\nu_2$ and $\nu_3$---which depends on their lifetimes---and on the flavor composition at the sources, which determines their initial abundance.  Thus, given that we do not know what are the neutrino lifetimes, we need to understand whether neutrino decay can alter flavor composition in a way that can be distinguished from the standard-oscillation expectation, regardless of the uncertainties in the flavor composition at the sources and the mixing parameters.  

Later, we explore this question by detailed calculation of the high-energy neutrino fluxes.  Here, we present qualitative insight.  We investigate the flavor composition at Earth in four extreme scenarios: \textit{(i)} no neutrino decay; \textit{(ii)} $\nu_2$ decay completely; \textit{(iii)} $\nu_3$ decay completely; \textit{(iv)} both $\nu_2$ and $\nu_3$ decay completely.  While the reality might lie in-between these scenarios, they illustrate the origin of our sensitivity to neutrino decay.  In each scenario, we generate the region of allowed flavor composition at Earth like before, by considering a generic flavor composition at the sources of $(f_{e, {\rm S}}, 1-f_{e, {\rm S}}, 0)$, varying $f_{e, {\rm S}}$ uniformly between $0$ and $1$, and varying the mixing parameters within their presently allowed ranges.

Figure~\ref{fig:flavor_triangle} shows the resulting regions.  Scenario \textit{(i)}, without decay, yields the standard, theoretically palatable region that we described earlier.  Scenario \textit{(ii)}, where only $\nu_2$ decay completely, yields a region that fully contains the standard region.  Because of the uncertainty in the flavor composition at the sources and in the mixing parameters, a clear separation of scenario \textit{(ii)} from scenario \textit{(i)} would be challenging even if a precise measurement of the flavor composition were available.  (Future improvements in the precision with which the mixing parameters are known would help~\cite{Song:2020nfh}.)  

Scenario \textit{(iii)}, where only $\nu_3$ decay completely, yields a region that only partially overlaps the standard region.  As a result, a precise measurement of the flavor composition could, in principle, separate \textit{(iii)} from \textit{(i)}.  In particular, if future measurements were to reveal that the flavor composition is either pion-beam or muon-damped, \figu{flavor_triangle} shows that this would disfavor scenario \textit{(iii)}.

Scenario \textit{(iv)}, where both $\nu_2$ and $\nu_3$ decay completely, yields a region of flavor composition given by the flavor content of the sole remaining eigenstate, $\nu_1$ (see the discussion above).  The region is small, separate from the standard region, and near the edge of the present-day region measured by IceCube~\cite{IceCube:2020fpi}.  This suggests that even present data might probe this scenario.  Indeed, later (Sec.~\ref{sec:bounds_diffuse}) we find that present observations of the diffuse neutrino flux can already exclude it at 95\%~C.L. 

\begin{figure}[t!]
 \includegraphics[width=\columnwidth]{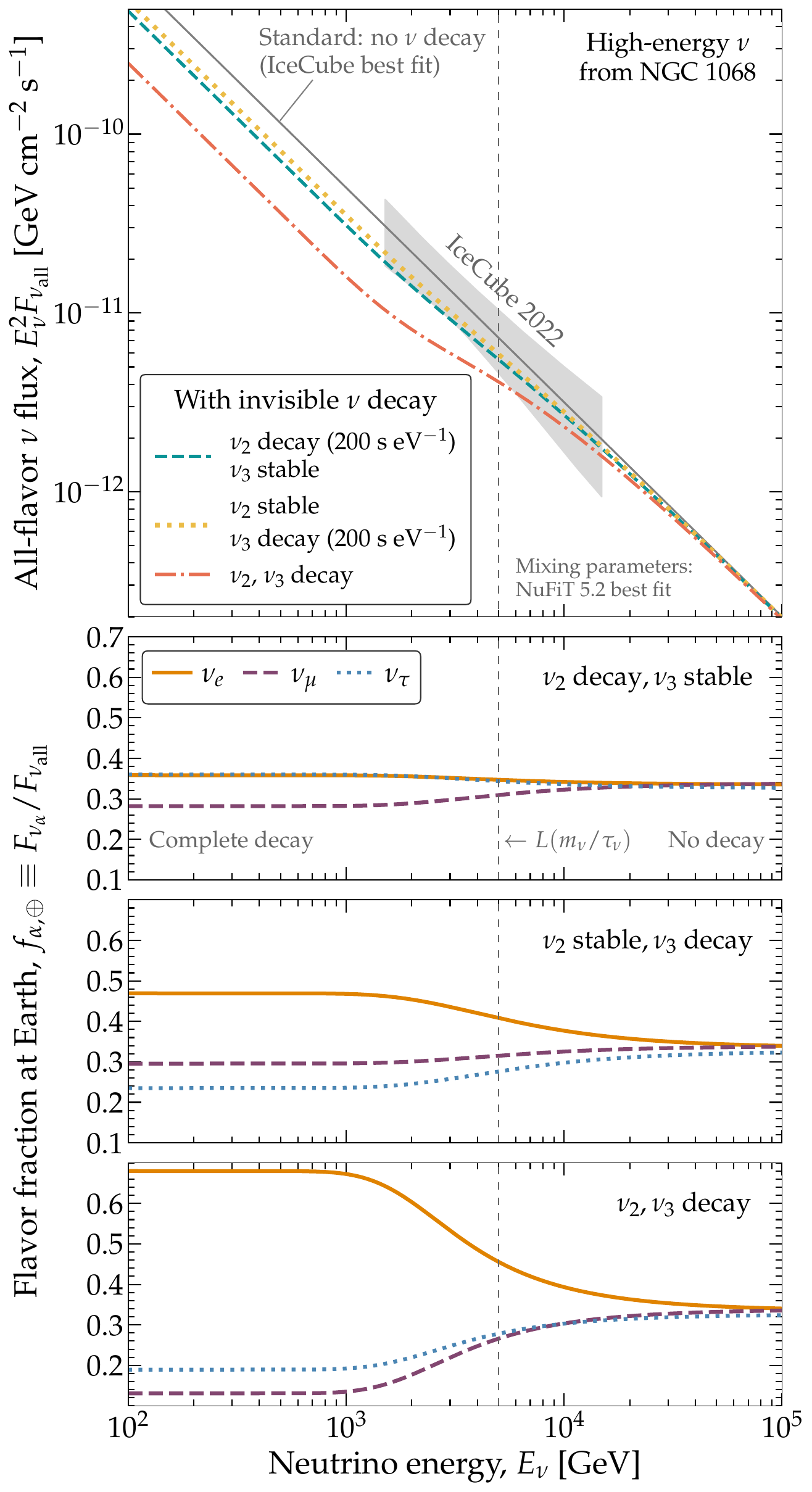}
 \caption{\textbf{\textit{Effect of neutrino decay on the high-energy neutrino flux from NGC 1068.}} In this figure, we fix the lifetime of decaying neutrinos to a common value of $\tau / m = 200$~s~eV$^{-1}$ as illustration; in our analysis, we vary the lifetime.  \textit{Top:} Neutrino flux at Earth, with and without the effect of $\nu_2$ decay, $\nu_3$ decay, and both.  IceCube measurements are from Ref.~\cite{IceCube:2022der}.  \textit{Bottom:} Neutrino flavor composition at Earth for the same cases, showing the approximate neutrino energy where neutrino decay becomes important. \textit{Neutrino decay induces spectral and flavor distortions within the IceCube energy window; the spectral distortions might disrupt the agreement with precision measurements of the neutrino flux.}}
 \label{fig:flux_ngc1068}
\end{figure}

\begin{figure}[t!]
 \includegraphics[width=\columnwidth]{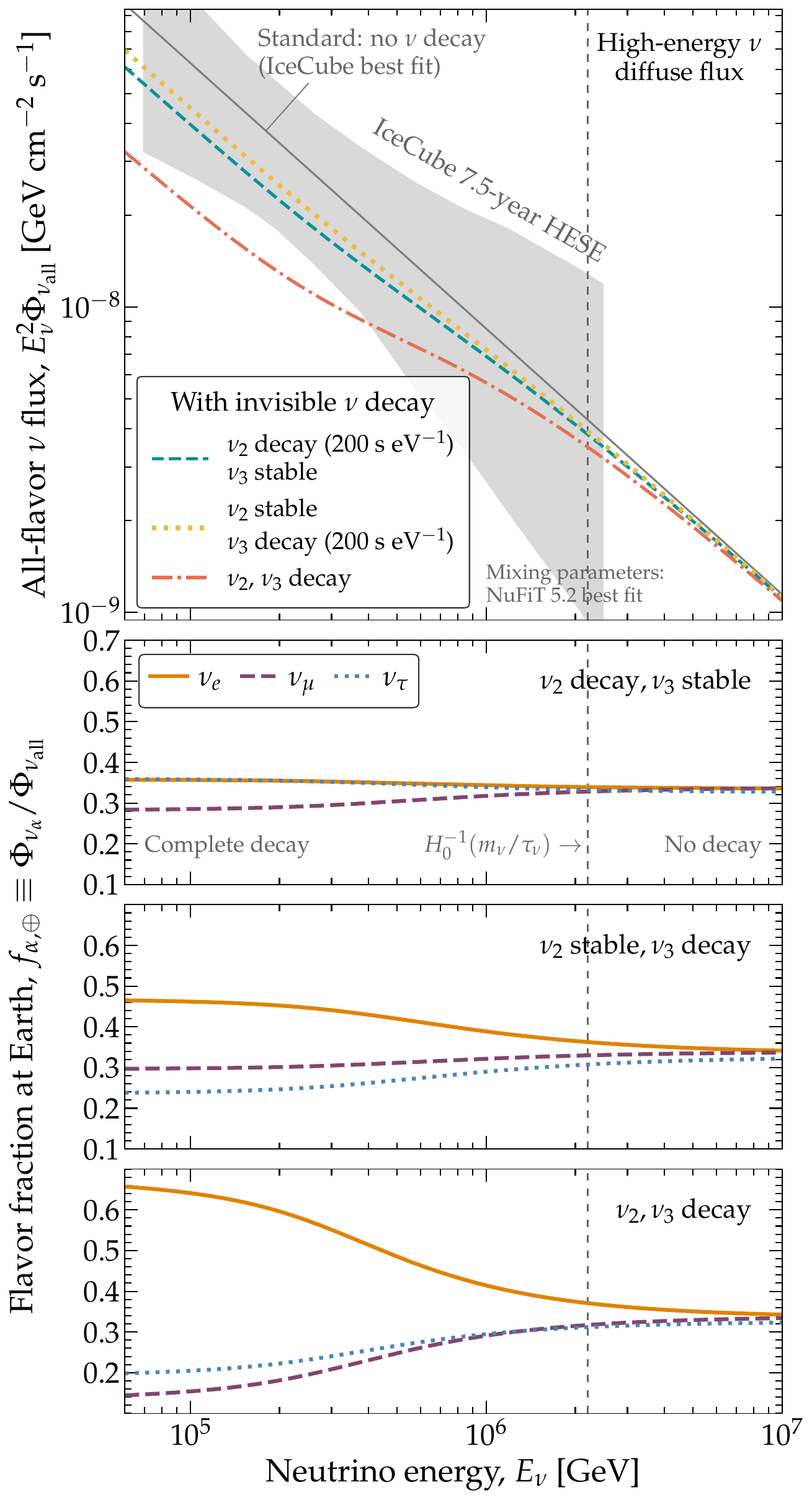}
 \caption{\textbf{\textit{Effect of neutrino decay on the high-energy diffuse neutrino flux.}} Similar to \figu{flux_ngc1068}, but for the diffuse flux of neutrinos detected by IceCube.  IceCube measurements are from Ref.~\cite{IceCube:2020wum}, based on HESE events. \textit{Both the neutrino energy distribution and flavor composition are altered by neutrino decay, providing a potential signature of its existence.}}
 \label{fig:flux_diffuse}
\end{figure}

In summary, while testing the decay of only $\nu_2$ via the flavor composition is challenging, testing the decay of only $\nu_3$, or of $\nu_2$ and $\nu_3$, has promising prospects, at least from the perspective of theory.  Later (Sec.~\ref{sec:bounds_diffuse}), we confirm the broad notions presented above via detailed analysis of neutrino observations.  However, the main limiting factor to testing neutrino decay via the flavor composition is not the uncertainty in the mixing parameters---which, nevertheless, we account for---but rather the fact that measuring the flavor composition of high-energy astrophysical neutrinos is difficult---which we also account for (Sec.~\ref{sec:bounds_diffuse-hese}); see also \Refe~\cite{Song:2020nfh}.  


\subsection{Neutrino decay in the energy spectrum}
\label{sec:nu_prop_with_decay-energy_spectrum}

In our analysis, we constrain neutrino decay using the flux of high-energy neutrinos from a point astrophysical source---specifically, NGC 1068---and from the diffuse flux of high-energy neutrinos.  The effect of neutrino decay on the energy spectra is similar in both.

\smallskip

\textbf{\textit{Point-source neutrino flux.---}}Under neutrino decay, the surviving flux $F_\alpha \equiv dN_\alpha / (dE_\nu dA dt)$ of $\nu_\alpha + \bar{\nu}_\alpha$ at Earth, per unit energy ($E_\nu$), area ($A$), and time ($t)$, emitted by a nearby source,\ie, one located at a distance $r \ll H_0^{-1} \approx 4$~Gpc, with $H_0$ the Hubble constant, is
\begin{equation}
 \label{equ:point_source_flux}
 F_\alpha(E_\nu)
 =
 \sum_{\beta,i} 
 \left\vert U_{\alpha i} \right\vert^2 
 \left\vert U_{\beta i} \right\vert^2 
 \frac{1}{4\pi r^2} 
 \frac{dN_\beta}{dE_\nu dt} 
 \exp\left[-\frac{m_i}{\tau_i} \frac{L}{E_\nu}\right] \;,
\end{equation}
where $dN_\beta/dE_\nu dt$ is the spectrum of $\nu_\beta$ emitted by the source, and the exponential factor reflects the attenuation due to decay (Sec.~\ref{sec:introduction}).  Later, to produce our results (Sec.~\ref{sec:bounds_ngc1068}), we use for the emitted spectrum either a power law in $E_\nu$, or one augmented by a high-energy cut-off.  

Figure~\ref{fig:flux_ngc1068} shows the neutrino flux from NGC 1068 computed using \equ{point_source_flux}, assuming for the emitted neutrino spectrum a power law whose normalization and spectral index are fixed to the best-fit values reported by the IceCube Collaboration~\cite{IceCube:2022der}.  For this figure only, as illustration, we assume that the flavor composition at Earth is $\left( \frac{1}{3}, \frac{1}{3}, \frac{1}{3} \right)_\oplus$ in the absence of decay.   We show the impact of the invisible decay of $\nu_2$ and $\nu_3$ on the flux on the all-flavor flux and on the flavor composition, for one choice of value of their lifetime, of 200~s~eV$^{-1}$, which makes the effects of decay appear within the energy window with which IceCube observes NGC 1068.  Later, when generating our results, we vary the lifetimes of $\nu_2$ and $\nu_3$.

Low-energy neutrinos decay faster because they are less boosted.  As a result, the all-flavor flux in \figu{flux_ngc1068} shows that the number of neutrinos drops at energies lower than about $r \cdot \mathrm{max}_i (m_i/\tau_i)$.  The drop is more pronounced if both $\nu_2$ and $\nu_3$ decay.  The flux transitions from no decay, at high energies, to complete decay, at low energies, over an energy window of about one order of magnitude.  The complete decay of $\nu_2$ or $\nu_3$ removes one third of the flux each; if both decay completely, only a third of the emitted flux survives.  If $\nu_2$ and $\nu_3$ have different lifetimes, their decay could introduce two drops in the flux, at different energies; see, \eg, \Refe~\cite{Abdullahi:2020rge}.  However, because in order for them to be detectable they have to occur within the IceCube energy window, which is rather narrow, and because the transition region is wide, it would be challenging to isolate them.

Figure~\ref{fig:flux_ngc1068} also shows how neutrino decay alters the flavor composition at Earth as a function of energy.  At high energies, the flavor composition is the one expected from standard oscillations in scenario \textit{(i)} in Sec.~\ref{sec:nu_prop_with_decay-flavor_comp}.  At low energies, if decay is complete, the flavor composition is akin to that expected from scenarios \textit{(ii)}, \textit{(iii)}, or \textit{(iv)}, depending on whether only $\nu_2$ decay, only $\nu_3$ decay, or both  decay.  In the transition region, the flavor composition is in-between scenarios \textit{(i)} and \textit{(ii)}, \textit{(iii)}, or \textit{(iv)}.  However, presently, the flavor composition of the neutrinos from NGC 1068 is experimentally inaccessible (Sec.~\ref{sec:nu_prop_with_decay-diffuse_vs_point}).

\smallskip

\textbf{\textit{Diffuse neutrino flux.---}}Although the identity, number, and distances to the astrophysical sources responsible for the diffuse neutrino flux are unknown (Sec.~\ref{sec:bounds_diffuse-flux}), we can estimate it by postulating a population of nondescript sources.  We assume that the diffuse flux is due to identical sources whose number density, $\rho_{\rm src}$, evolves with redshift, each source emitting neutrinos with the same spectrum.

Unlike the flux from a nearby point source, the sources that make up the diffuse flux are likely located farther away, at redshifts of $z = 1$ or above, \ie, at hundreds of Mpc to a few Gpc.  Therefore, we need to account for the effect of the cosmological expansion on the energies of the neutrinos as they propagate to Earth, which changes the size of their Lorentz boost and, as a result, their susceptibility to decay; see \Refe~\cite{Baerwald:2012kc} for details.  

Under decay, the surviving flux $\Phi_\alpha \equiv dN_\beta / (dE_\nu dA dt d\Omega)$ of $\nu_\alpha + \bar{\nu}_\alpha$ at Earth, per unit solid angle ($\Omega$), is the sum of contributions from sources across all redshifts~\cite{Baerwald:2012kc, Bustamante:2016ciw}, \ie,
\begin{eqnarray}
 \label{equ:diffuse_integration}
 && 
 \Phi_\alpha(E_\nu)
 =
 \sum_{\beta,i} 
 \left\vert U_{\alpha i} \right\vert^2 
 \left\vert U_{\beta i} \right\vert^2 
 \int_0^{\infty}
 \frac{dz}{H(z)} 
 \rho_{\rm src}(z)
 \\  
 \nonumber 
 && 
 \times
 \frac{dN_\beta \left[E_\nu(1+z)\right] }{dE_\nu dt}
 \exp 
 \left[
 -\frac{\tau_i}{m_i}
 \frac{1}{E_\nu}
 \int_0^z \frac{dz'}{H(z')(1+z')^2}
 \right] \;,
\end{eqnarray}
where $H(z) = H_0 \left[ \Omega_\Lambda + (1+z)^3 \Omega_m \right]^{1/2}$ is the Hubble parameter.  We assume a $\Lambda$CDM cosmology with $H_0 = 64.7$~km~s$^{-1}$~Mpc$^{-1}$, and adimensional energy density parameters $\Omega_\Lambda = 0.685$ and $\Omega_m = 0.315$~\cite{ParticleDataGroup:2022pth}.  Effectively, contributions from sources at $z \gtrsim 4$ are negligible because they are rare and distant.  

Figure~\ref{fig:flux_diffuse} shows the diffuse neutrino flux computed using \equ{diffuse_integration}, assuming for the emitted neutrino spectrum a power law whose normalization and spectral index are fixed to the best-fit values reported by the IceCube Collaboration in their analysis of the 7.5-year sample of High-Energy Starting Events (HESE)~\cite{IceCube:2020wum}.   As in \figu{flux_ngc1068}, for this figure we assume that the flavor composition at Earth is $\left( \frac{1}{3}, \frac{1}{3}, \frac{1}{3} \right)_\oplus$ in the absence of decay.  We assume that $\rho_{\rm src}$ follows the star formation rate~\cite{Hopkins:2006bw}, which places most sources around $z = 1$ (more on this later).

Similarly to the point-source neutrino flux, the invisible decay of $\nu_2$ and $\nu_3$ affects the all-flavor diffuse flux and its flavor composition.  Compared to the point-source flux, however, the transition from no decay to complete decay in the diffuse flux is smoothed out, since the sources are located at different redshifts and what we see at Earth is the superposition of the transitions in the individual neutrino spectra they emit.  Different from the point-source flux, we do have the capacity at present to measure the flavor composition of the diffuse flux (Sec.~\ref{sec:nu_prop_with_decay-diffuse_vs_point}).


\subsection{Point-source {\it vs.} diffuse neutrino fluxes}
\label{sec:nu_prop_with_decay-diffuse_vs_point}

The neutrinos from NGC 1068 and the diffuse neutrino flux offer complementary sensitivity to neutrino decay:
\begin{description}
 \item[Different energy ranges] Neutrinos detected from NGC 1068~\cite{IceCube:2022der} have lower energies, of roughly 1--10~TeV, than the neutrinos that make up the diffuse flux~\cite{IceCube:2020wum}, of roughly 60~TeV--10~PeV.  As a result, neutrinos from NGC 1068 have a weaker Lorentz boost (Sec~\ref{sec:introduction}) and are therefore sensitive to factor-of-10 longer lifetimes.
 \item[Different observables] Because neutrinos from NGC 1068 have so far been detected only via $\nu_\mu$-initiated muon tracks, with them we can look for the effects of decay solely on the neutrino energy spectrum.  In contrast, because the diffuse neutrino flux has been detected using HESE events, which are differently sensitive to all flavors (more on this later), with it we can look for the effects of decay jointly on the energy spectrum and flavor composition.
 \item[Comparable uncertainties] Presently, the number of events detected by IceCube from NGC 1068 and from the diffuse flux are roughly comparable (though the former suffers from a higher background).  The relative uncertainties with which the flux normalization and spectral index are measured (\cf~\Refes~\cite{IceCube:2020wum} and \cite{IceCube:2022der}) are comparable between the two fluxes, assuming a power-law neutrino flux.
\end{description}

There is one more subtle difference.  The neutrinos from NGC 1068 demonstrably originate from a single source, the distance to which is known.  Their energy spectrum is due only to that one source, all neutrinos presumably created in the same processes and under the same physical conditions.  Taken at face value, these facts should seemingly reduce the impact of astrophysical uncertainties on the sensitivity to neutrino decay that can be garnered from NGC 1068.  In practice, this is not so.  

The reason is two-fold.  First, because for NGC 1068 we can only look for signs of decay in the all-flavor neutrino energy spectrum---and not also in the flavor composition---the sensitivity is limited to spotting a drop in the flux normalization of, at most, a factor of 3 (Sec.~\ref{sec:nu_prop_with_decay-energy_spectrum}), which is within the uncertainty with which IceCube has measured the flux so far (\figu{flux_ngc1068}).  Second, there are large theory unknowns in the prediction of the flux, all viable given present experimental uncertainties, that further cloud potential evidence of decay.  In our analysis below, we account for both limitations, show how they weaken the sensitivity to neutrino decay significantly, and how in the future they might be mitigated.

In contrast, the diffuse neutrino flux is made up of contributions from an unknown number of unidentified neutrino sources whose distance distribution is also unknown, each one likely producing neutrinos via mechanisms that are at least somewhat different.  Taken at face value, these astrophysical uncertainties should render the sensitivity to neutrino decay from the diffuse flux null.  In practice, this is not so.

The reason, again, is two-fold.  First, even though there is likely an unquantified level of variation in the neutrino spectra and flavor composition emitted by the sources, the diffuse flux that we measure at Earth is their population-averaged aggregate, where presumably this variation has been smoothed out.  Second---and more importantly---with the diffuse flux we can look for signs of decay in the flavor composition.  Given that the measurement uncertainty of the diffuse flux normalization is comparable to that of NGC 1068---where it is large enough to hide potential signs of decay---it is the flavor composition that delivers the key advantage needed to test neutrino decay.

Below, in Secs.~\ref{sec:bounds_ngc1068} and \ref{sec:bounds_diffuse}, we show how the above perspectives are evident in our results.


\section{Neutrino lifetime bounds from NGC 1068}
\label{sec:bounds_ngc1068}

We compare the observation of high-energy neutrinos from NGC 1068 against theoretical expectations of the neutrino flux, with and without decay.  First we review the salient features of models of astrophysical neutrino production in NGC 1068 and discuss the choices we make for it in our analysis.  Then we introduce the statistical procedure we follow to obtain constraints on neutrino decay, and show our results.


\subsection{Astrophysical neutrinos from NGC 1068}
\label{sec:bounds_ngc1068-flux}

Recently, the IceCube Collaboration reported the discovery of an excess of $79_{-20}^{+22}$ neutrinos with about 1--10~TeV from the direction of the active galactic nucleus (AGN) of the Seyfert galaxy NGC~1068~\cite{IceCube:2019cia,IceCube:2022der}, which constitutes an excess over the atmospheric and cosmic background expectations with a statistical significance of $4.2\sigma$.  This makes NGC 1068 the first candidate steady-state source of high-energy astrophysical neutrinos.

\smallskip

\textbf{\textit{Neutrino production in NGC 1068.---}}Although NGC 1068 is likely a high-energy neutrino source, to date it is unclear how and where inside it the neutrinos are made.  There are, however, well-motivated hypotheses.

The absence of gamma rays at energies comparable to those of the neutrinos~\cite{MAGIC:2019fvw} disfavors neutrino production in the outer regions of the galaxy, such as the jet~\cite{Lenain:2010kc}, the circumnuclear starburst region~\cite{Yoast-Hull:2013qfa, Ambrosone:2021aaw, Eichmann:2022lxh}, a large-scale AGN-driven outflow~\cite{Lamastra:2016axo}, or an ultra-fast outflow~\cite{Peretti:2023xqk}.  

This suggests that neutrino production occurs in an inner, gamma ray-opaque region~\cite{Murase:2022dog}, which could be the hot AGN corona, where cosmic-ray protons ($p$) and nuclei could undergo neutrino-producing photohadronic ($p \gamma$) interactions on a dense X-ray field~\cite{Inoue:2019yfs, Murase:2019vdl, Kheirandish:2021wkm, Mbarek:2023yeq, Fiorillo:2023dts}.  Alternatively, neutrino production in inner regions different from the corona, \eg, the inner circumnuclear region~\cite{Fang:2023vdg}, could be dominated by cosmic-ray interactions on surrounding matter ($pp$).  In either case, the interactions produce high-energy pions that, upon decaying, produce high-energy neutrinos (Sec.~\ref{sec:nu_prop_with_decay-flavor_comp}).  The specific production channel, and also the cosmic-ray acceleration mechanism, impact the energy spectrum of the neutrinos.  Later, we show that the uncertainty in these predictions impacts our sensitivity to neutrino decay.

Already today, neutrino and gamma-ray observations of NGC 1068 provide valuable model-independent constraints on the neutrino spectrum.  The IceCube observations~\cite{IceCube:2022der} show that the neutrino energy flux $E_\nu^2 \Phi_\nu$ falls with energy above $1.5$~TeV.  Fitting the observations with a power law, $\Phi_\nu\propto E_\nu^{-\gamma}$, yields a best-fit spectral index $\gamma= 3.2$~\cite{IceCube:2022der}; see \figu{flux_ngc1068}.  However, such a soft energy spectrum cannot be extended to arbitrarily low energies, since the electromagnetic emission co-produced with the neutrinos would exceed the bolometric luminosity of the source~\cite{Murase:2022dog}. Thus, a consistent picture requires a relatively hard neutrino spectrum below about $1$~TeV, with $\gamma < 2$, and a softer spectrum at higher energies. 

\smallskip

\textbf{\textit{Neutrino flux models.---}}Broadly stated, the above requirements can be met in two ways.  First, it can be obtained by cosmic rays in the source reaching energy equipartition with the magnetic field~\cite{Fiorillo:2023dts}.  This yields a soft power-law neutrino spectrum with $\gamma \approx 3$ across the entire energy range.  Second, it can be obtained in models of coronal emission where a change in the spectral index is due to cosmic rays reaching their maximum energy~\cite{Inoue:2019yfs, Murase:2019vdl, Kheirandish:2021wkm, Mbarek:2023yeq}.  This yields a neutrino spectrum above 1~TeV that is suppressed by an exponential cut-off $\propto e^{-E_\nu/E_{\nu,{\rm cut}}}$, with $E_{\nu, {\rm cut}}$ a characteristic cut-off energy (alternatively, it could also yield a broken power-law spectrum~\cite{Fiorillo:2023dts}).

In our work, we are only interested in neutrinos above $1$~TeV, where the IceCube observations are presently available.  Thus, to reflect the above possibilities, we consider two alternative parametrizations for the energy spectrum: a power law (PL),
\begin{equation}
 E_\nu^2 \Phi_\nu
 =
 (E^2_{\nu} \Phi_\nu)_0 
 \left(
 \frac{E_\nu}{1~{\rm TeV}}
 \right)^{2-\gamma} \;,
\end{equation}
where the two free parameters are the flux normalization, $(E^2_{\nu} \Phi_\nu)_0$, and the spectral index, $\gamma$, and a power law augmented by a cut-off (PC),
\begin{equation}
 E_\nu^2 \Phi_\nu
 =
 (E^2_{\nu} \Phi_\nu)_0 
 \left(
 \frac{E_\nu}{1~{\rm TeV}}
 \right)^{2-\gamma}
 \exp\left(-\frac{E_\nu}{E_{\nu,\mathrm{cut}}}\right) \;,
\end{equation}
where the three free parameters are $(E^2_{\nu} \Phi_\nu)_0$, $\gamma$, and $E_{\nu,\mathrm{cut}}$.  Later, we allow the values of these parameters to float. The PC model reproduces the PL model when $E_{\nu,\mathrm{cut}} \gg 15$~TeV, above the range measured by IceCube.

The degeneracy between the PL and PC models reflects the fact that, in the absence of neutrino decay, present-day IceCube observations are described well either by a soft-spectrum PL model, or by a PC model with little sensitivity to the value of the spectral index at energies below the cut-off.  In the presence of neutrino decay, the degeneracy allows for more freedom in fitting the observations, and thus yields more conservative bounds on neutrino decay.   Our two scenarios, PL and PC, encompass the theoretical possibilities proposed so far to explain the neutrino signal from NGC 1068. 

\smallskip

\textbf{\textit{Detecting neutrinos from NGC 1068.---}}Presently, IceCube observes neutrinos from NGC 1068 exclusively via \textit{tracks}.  These are events where a $\nu_\mu$ interacts with a nucleon in the ice and generates a high-energy muon that produces a km-scale track of Cherenkov light as it propagates.  The interaction of a $\nu_\tau$, and the subsequent decay of the final-state tau, also produces a high-energy muon 17\% of the time~\cite{ParticleDataGroup:2022pth}, but this contribution cannot be isolated from that of $\nu_\mu$.  We expand on neutrino detection in IceCube in Sec.~\ref{sec:bounds_diffuse-hese}.  Because the direction tracks can be reconstructed accurately, typically to less than $1^\circ$, they are regularly used to look for point neutrino sources. 

The IceCube Collaboration uses a combination of \textit{starting} and through-going \textit{tracks}---where the muon is created inside or outside the detector volume, respectively---to infer the energy spectrum of $\nu_\mu$ from NGC 1068 (with a small contamination from $\nu_\tau$)~\cite{IceCube:2022der}.  In through-going tracks,  only a segment of the muon track crosses the detector.  This makes inferring the original neutrino energy from the energy deposited by the muon in the ice especially challenging.  To make our analysis realistic, when computing event rates (Sec.~\ref{sec:bounds_ngc1068-event_rates}), we also combine starting and through-going tracks and account for the uncertainty in the energy reconstruction.

By using exclusively tracks, IceCube is presently unable to measure the flavor composition of the neutrino flux from NGC 1068.  This limits our search for neutrino decay to its effects on the energy spectrum (Sec.~\ref{sec:nu_prop_with_decay-energy_spectrum}).  Later, we show that this lack of flavor information greatly weakens our sensitivity to neutrino decay.  

Yet, although flavor measurements are unavailable, the changes to the $\nu_\mu$ spectrum induced by neutrino decay (Sec.~\ref{sec:nu_prop_with_decay-diffuse_vs_point}) do depend on the flavor composition with which the neutrinos are emitted.  Below, we show how.

\smallskip

\textbf{\textit{Flavor composition.---}}Most models of neutrino production in NGC 1068 predict a pion-beam flavor composition (Sec.~\ref{sec:nu_prop_with_decay-flavor_comp}), \ie, $\left( \frac{1}{3}, \frac{2}{3}, 0 \right)_{\rm S}$.  This is a robust prediction.  Even in the extreme case where the AGN corona is very compact and the energy densities of the magnetic field and X-rays in it are equal, the magnetic field intensity in the region would be about $10^4$~G~\cite{Fiorillo:2023dts}, which would induce a transition to the muon-damped regime only at energies above $100$~TeV~\cite{Winter:2013cla, Bustamante:2020bxp, Fiorillo:2021hty}, well beyond the IceCube range.  Hence, we consider as our baseline scenario a pion-beam flavor composition when comparing our predictions to observations, either real or projected.

In addition, we also test a generic scenario with no $\nu_\tau$ production and with the form $(f_{e,{\rm S}}, 1-f_{e,{\rm S}}, 0)$ (Sec.~\ref{sec:nu_prop_with_decay-flavor_comp}), where we vary the $\nu_e$ fraction, $f_{e,{\rm S}} \in [0,0.5]$ when comparing our predictions to observations.  We avoid  values of $f_{e, {\rm S}} > 0.5$ because they correspond to a neutron-beam composition (Sec.~\ref{sec:nu_prop_with_decay-flavor_comp}), which is unlikely.  If the TeV neutrinos observed by IceCube were produced as $\bar{\nu}_e$ in the decay of neutrons created in $pp$ or $p\gamma$ interactions, we would expect associated production of PeV neutrinos from the decay of pions produced in the same interactions, which we have not seen.  This generic form of the flavor composition allows us to smoothly shift between the pion-beam and muon-damped benchmarks. 

\smallskip

\textbf{\textit{Flux shape and normalization}}.---In our baseline scenario, we consider the normalization of the neutrino flux and the shape of its energy spectrum from NGC 1068 ($\Phi_0$ and $\gamma$ below, respectively) as free parameters, with no prior information on them, \ie, without any associated pull terms in our test statistic (Sec.~\ref{sec:bounds_ngc1068-stat}).  This choice reflects the challenge of our present situation, where, barring adopting an unduly precise prescription of the neutrino flux at emission time, we are confined to looking for signs of decay in the flux at Earth while having little to no knowledge of what it was at emission time, before it could have been affected by neutrino decay.

In the future, this situation could improve by using X-ray observations of NGC 1068.  The processes responsible for TeV neutrino production also produce TeV gamma rays that, upon interacting with their environment, cascade down to lower energies and are expected to emerge from the source with MeV-scale energies.  Future observations from planned experiments like AMEGO-X~\cite{Caputo:2022xpx} and e-ASTROGAM~\cite{e-ASTROGAM:2016bph}, could detect these photons.  On the basis of the connection between neutrino and gamma-ray energies, these observations would enable us to estimate the energy in neutrinos emitted by the source, before their decay could possibly affect them.  Such estimate would require a detailed understanding of the cascade process and the source environment.  For now, it remains as an optimistic future possibility.  We entertain it only by proxy by considering an alternative scenario (``tight prior'' in Table~\ref{tab:ngc_1068}) where we impose a Gaussian prior on the total energy contained in the neutrino flux in the absence of decay, $E_{\nu, {\rm tot}}$, centered at the value of the energy integral of the neutrino flux reported by IceCube~\cite{IceCube:2022der}, with a wide standard deviation of half a decade.

\begingroup
\begin{table*}[t!]
 \begin{ruledtabular} 
  \caption{\label{tab:ngc_1068}\textbf{\textit{Constraints on the neutrino lifetime from the flux of high-energy astrophysical neutrinos from NGC 1068.}}  Constraints are on the lifetimes of $\nu_2$ and $\nu_3$, computed from projected, cumulative detection of through-going and starting muon tracks by the combination, in PLE$\nu$M~\cite{Schumacher:2021hhm}, of neutrino telescopes Baikal-GVD, IceCube, IceCube-Gen2, KM3NeT, and P-ONE by the year 2035.  No constraints can be placed using present-day neutrino observations of NGC 1068 by IceCube alone~\cite{IceCube:2022der}.  The constraints shown are after profiling over all other model parameters, for our eight scenarios of the shape of the neutrino energy spectrum [power law (PL) and power law with high-energy exponential cut-off (PC)], the use of a prior on the neutrino flux normalization (none and tight), and the treatment of the flux composition at the sources, $f_{\rm S}$ (freely floating and fixed to the pion-beam composition).  Two-dimensional constraints are preferred because they depict the correlation between $\tau_2/m_2$ and $\tau_3/m_3$, which is not captured by the one-dimensional constraints.  Blank entries, marked with ``$\cdots$'', represent scenarios where no constraint can be placed.  See \figu{1d_bounds_summary} for a graphical comparison of all one-dimensional constraints and Secs.~\ref{sec:nu_prop_with_decay} and \ref{sec:bounds_ngc1068} for details.}
  \centering
  \renewcommand{\arraystretch}{1.7}
  \begin{tabular}{cccccccccc}
   \multirow{4}{*}{\makecell{Shape of the \\ neutrino \\ energy \\ spectrum}}        &
   \multirow{4}{*}{\makecell{Prior on the \\ neutrino \\ flux \\ normalization}}     &
   \multicolumn{8}{c}{\makecell{Neutrino lifetime disfavored at 95\%~C.L., $\tau_i/m_i$ [s~eV$^{-1}$] \\ (projected, year 2035, multiple detectors, tracks)}} 
   \\
   \cline{3-10}
   &
   &
   \multicolumn{4}{c}{$f_{\rm S}$ free}                       &
   \multicolumn{4}{c}{$f_{\rm S}$ fixed to $\pi$ decay}       \\
   \cline{3-6}
   \cline{7-10}
   &
   &
   \multirow{2}{*}{\makecell{$\frac{\tau_3}{m_3}$~{\it vs.}~$\frac{\tau_2}{m_2}$ \\[0.5em] (preferred)}}   &   
   \multirow{2}{*}{$\frac{\tau_2}{m_2}$\footnote{\label{fnote1}One-dimensional lower limit after profiling over $\tau_3/m_3$.  See also \figu{1d_bounds_summary}.}}    &
   \multirow{2}{*}{$\frac{\tau_3}{m_3}$\footnote{\label{fnote2}One-dimensional lower limit after profiling over $\tau_2/m_2$.  See also \figu{1d_bounds_summary}.}}    &
   \multirow{2}{*}{$\frac{\tau_2}{m_2} = \frac{\tau_3}{m_3}$\footnote{\label{fnote3}One-dimensional lower limit obtained assuming $\tau_2/m_2 = \tau_3/m_3$.  See also \figu{1d_bounds_summary}.}}                                                 &   
   \multirow{2}{*}{\makecell{$\frac{\tau_3}{m_3}$~{\it vs.}~$\frac{\tau_2}{m_2}$ \\[0.5em] (preferred)}}   &   
   \multirow{2}{*}{$\frac{\tau_2}{m_2}$}                                       &
   \multirow{2}{*}{$\frac{\tau_3}{m_3}$}                                       &
   \multirow{2}{*}{$\frac{\tau_2}{m_2} = \frac{\tau_3}{m_3}$}                  \\
   \\
   \hline
   PL                                                         &
   None                                                       &
   \figu{bounds_ngc1068}                                      &
   $\cdots$                                                   &
   $\cdots$                                                   &
   $[19.63, 6382.63]$                                         &
   \figu{bounds_ngc1068}                                      &
   $\cdots$                                                   &
   $\cdots$                                                   &
   $[11.32, 7430.19]$                                         \\
   PL                                                         &
   Tight\footnote{\label{fnote4}The tight prior on the total energy in neutrinos, $E_{\nu, {\rm tot}}$, is a Gaussian centered on the energy integral of the present-day IceCube measurement of the neutrino flux from NGC 1068~\cite{IceCube:2022der}, between $10^4$ and $10^7$~GeV, and has a standard deviation of half a decade.}                      &
   \figu{bounds_ngc1068}                                      &
   $\cdots$                                                   &
   $\cdots$                                                   &
   $[12.56, 6839.12]$                                         &
   \figu{bounds_ngc1068}                                      &
   $\cdots$                                                   &
   $\cdots$                                                   &
   $[4.94, 7961.59]$                                          \\
   PC                                                         &
   None                                                       &
   \figu{bounds_ngc1068}                                      &
   $\cdots$                                                   &
   $\cdots$                                                   &
   $[483.05, 1945.36]$                                        &
   \figu{bounds_ngc1068}                                      &
   $\cdots$                                                   &
   $\cdots$                                                   &
   $[365.59, 2371.37]$                                        \\  
   PC                                                         &
   Tight                                                      &
   \figu{bounds_ngc1068}                                      &
   $\cdots$                                                   &
   $\cdots$                                                   &
   $[392.64, 2023.02]$                                        &
   \figu{bounds_ngc1068}                                      &
   $\cdots$                                                   &
   $\cdots$                                                   &
   $[283.14, 2415.46]$                                        \\  
  \end{tabular}
 \end{ruledtabular}
\end{table*}
\endgroup

Table~\ref{tab:ngc_1068} summarizes the eight scenarios that we explore using the neutrino flux from NGC 1068: two alternative shapes of the energy spectrum (PL and PC), two choices for the treatment of the prior on the flux normalization (none and tight), and two choices for the treatment of the flavor composition at the sources (freely floating and fixed to pion-beam).  For each scenario, we compute constraints on $\tau_2/m_2$ and $\tau_3/m_3$.


\subsection{Computing event rates}
\label{sec:bounds_ngc1068-event_rates}

\textbf{\textit{Present and projected observations.---}}So far, IceCube is the only telescope to identify high-energy neutrinos from NGC 1068.  We find later (Sec.~\ref{sec:bounds_ngc1068-results}) that, in given the large astrophysical unknowns, the present number of IceCube events from NGC 1068---about 80---is insufficient to search for neutrino decay.  Fortunately, in the near future, upcoming large-scale neutrino telescopes, currently under construction and planning~\cite{MammenAbraham:2022xoc, Ackermann:2022rqc, Guepin:2022qpl}, will boost the detection rate of high-energy neutrinos, including those coming from NGC 1068.

Thus, we perform two separate searches for neutrino decay in the neutrinos from NGC 1068.  First, we use the present-day data reported by the IceCube Collaboration in \Refe~\cite{IceCube:2022der}.  Second, we use the combined projected data collected by IceCube and by upcoming ice- and water-Cherenkov neutrino telescopes Baikal-GVD~\cite{Baikal-GVD:2023beh}, IceCube-Gen2~\cite{IceCube-Gen2:2020qha}---the envisioned high-energy upgrade of IceCube---KM3NeT~\cite{KM3Net:2016zxf}, and P-ONE~\cite{P-ONE:2020ljt}.  In all cases, we consider exclusively track events (see, however, Sec.~\ref{sec:bounds_ngc1068-future} for future improvements).

\smallskip

\textbf{\textit{Using PLE$\nu$M to compute detection.---}}Presently, there are no publicly available resources provided by the IceCube Collaboration to compute with high detail the response of the detector to neutrinos from NGC 1068.  [This is different when using HESE events to measure the diffuse neutrino flux (Sec.~\ref{sec:bounds_diffuse}).]  Hence, we compute event rates using the third-party public code~\cite{PlenumCode} associated to the Planetary Neutrino Monitoring System (PLE$\nu$M) network~\cite{Schumacher:2021hhm}.  PLE$\nu$M is a framework for boosting the information that can be gleamed from the detection of high-energy astrophysical neutrinos by present and future neutrino telescopes, by analyzing their observations within a common computational structure.

The PLE$\nu$M code reproduces the present-day IceCube response to neutrinos from NGC 1068~\cite{IceCube:2022der} and, based on it, forecasts the capabilities of future telescopes.  It computes neutrino detection by accounting for the effective areas, response functions, and energy and directional resolution of the different detectors listed above.  Event rates are computed in terms of the reconstructed muon energy, \ie, the energy of the muon as estimated from the light deposited in the detector by its track.  Doing so accounts for the relation between the energy of the parent neutrino, the energy of the  muon that it produces, and the experimentally measurable proxy of its energy.  

The relation between the neutrino energy and the true muon energy reflects the inelasticity distribution in charged-current neutrino-nucleon interactions, \ie, the distribution of the fraction of neutrino energy that is transferred to the final-state muon.  The relation between the true muon energy and reconstructed muon energy is obtained by means of detailed detector simulations.  To produce our results, we calibrate it to reproduce the recent improvements that led to the detection of neutrinos from NGC 1068 (Fig.~S5 in \Refe~\cite{IceCube:2022der}).  Because the conversion from neutrino energy to reconstructed muon energy smears any features that might exist in the neutrino energy spectrum, such as those induced by neutrino decay (\figu{flux_ngc1068}), it muddles their identification.  

We bin events in reconstructed muon energy and angular distance from the position of NGC 1068, similarly to Fig.~2 in \Refe~\cite{IceCube:2022der}.  We add to the signal from NGC 1068 the background of tracks due to atmospheric neutrinos and muons, computed, as in \Refe~\cite{IceCube:2022der}, using the \textsc{Sibyll~2.3c} hadronic model for particle interactions and the primary cosmic-ray flux from \Refe~\cite{Gaisser:2011klf}.


\subsection{Statistical analysis}
\label{sec:bounds_ngc1068-stat}

As discussed above (Sec.~\ref{sec:bounds_ngc1068-flux}), we consider two scenarios for the energy spectrum of high-energy neutrinos from NGC 1068.  In the PL scenario, the spectrum is a power law, determined by the set of astrophysical parameters $\boldsymbol{\theta}_{\rm ast}^{\rm PL} = \left\{ \Phi_0, \gamma, f_{e, {\rm S}} \right\}$, the mixing parameters $\boldsymbol{\theta}_{\rm mix} = \left\{ \theta_{12}, \theta_{23}, \theta_{13}, \delta_{\rm CP} \right\}$, and the decay parameters $\boldsymbol{\theta}_{\rm{dec}}=\left\{m_2/\tau_2, m_3/\tau_3\right\}$.  For the PC model, the set of astrophysical parameters is enlarged to account for the cut-off energy, \ie, $\boldsymbol{\theta}_{\rm ast}^{\rm PC} = \left\{ \Phi_0, \gamma, f_{e, {\rm S}}, E_{\nu, {\rm cut}} \right\}$.

For each choice of the above parameters, we use the PLE$\nu$M code (Sec.~\ref{sec:bounds_ngc1068-event_rates}) to generate the associated event sample, binned in reconstructed muon energy and angular distance from the position of NGC 1068.  Similarly to \Refe~\cite{IceCube:2022der}, we use $N_{\hat{E}_\mu} = 140$ bins in energy, evenly spaced in logarithmic scale between $10^2$ and $10^9$~GeV, and $N_{\Delta \psi} = 225$ bins in angular separation, evenly spaced between 0 and 9 sr.  We denote by $\mu_{ij}^{\nu, {\rm ast}}(\boldsymbol{\theta}_{\rm ast}, \boldsymbol{\theta}_{\rm mix}, \boldsymbol{\theta}_{\rm{dec}}, T)$ the mean expected number of tracks induced by the astrophysical neutrino flux from NGC 1068 in the $i$-th energy bin and $j$-th angular bin, computed for the choice of parameters $\boldsymbol{\theta}_{\rm{ast}}$ (either $\boldsymbol{\theta}_{\rm{ast}}^{\rm PL}$ or $\boldsymbol{\theta}_{\rm{ast}}^{\rm PC}$), $\boldsymbol{\theta}_{\rm{mix}}$, and $\boldsymbol{\theta}_{\rm{dec}}$, and for an exposure time $T$.

To account for the contamination of tracks due to the background of atmospheric neutrinos and muons, first, we compute the baseline background event distribution, $N_{ij}^\mu(T)$.  Then, we keep its spectral and angular shape  fixed, but allow for its size to be rescaled by floating its normalization constant, $\mathcal{N}^{\mu}$.  Thus, the number of background tracks is  $\mu_{ij}^{\rm atm}(\mathcal{N}^{\mu}, T) = \mathcal{N}^{\mu} N_{ij}^\mu(T)$.

Altogether, the mean number of tracks in each bin, of astrophysical and atmospheric origin, is
\begin{eqnarray}
 \label{equ:event_rate_ngc1068}
 &&
 \mu_{ij}
 (\boldsymbol{\theta}_{\rm ast},
 \boldsymbol{\theta}_{\rm mix},
 \boldsymbol{\theta}_{\rm{dec}},
 \mathcal{N}^{\mu},
 T)
 \nonumber \\
 &&
 \qquad
 =
 \mu_{ij}^{\rm ast}
 (\boldsymbol{\theta}_{\rm ast},
 \boldsymbol{\theta}_{\rm mix},
 \boldsymbol{\theta}_{\rm{dec}},
 T)
 +
 \mu_{ij}^{\mu}
 (\mathcal{N}^{\mu},
 T) \;.
\end{eqnarray}

As stated above, we perform two analyses: first, using the present-day IceCube observations of NGC 1068 and, second, using projections.  We describe them below.

\smallskip

\textbf{\textit{Using present-day data.---}}We compare the expected number of tracks, $\mu_{ij}$, against the sample of events reported by IceCube~\cite{IceCube:2022der}, in each energy and direction bin, $n_{ij}$. To compare them, we use a binned Poisson likelihood function that spans all bins, \ie,
\begin{eqnarray}
 \label{equ:likelihood_ngc1068}
 &&
 \ln
 \mathcal{L}
 (\boldsymbol\theta_{\rm ast},
 \boldsymbol{\theta}_{\rm{mix}},
 \boldsymbol\theta_{\rm dec}, 
 \mathcal{N}^\mu,
 T)
 \\
 &&
 \qquad
 =
 \sum_{i=1}^{N_{\hat{E}_\mu}}
 \sum_{j=1}^{N_{\Delta\psi}}
 \ln
 \mathcal{L}_{ij,{\rm t}}
 (\boldsymbol\theta_{\rm ast},
 \boldsymbol{\theta}_{\rm{mix}},
 \boldsymbol\theta_{\rm dec}, 
 \mathcal{N}^\mu,
 T) 
 \nonumber
 \\
 &&
 \qquad \quad
 +
 \frac{\chi^2(\boldsymbol{\theta}_{\rm{mix}})}{2}
 +
 \frac{\chi^2(\mathcal{N}^\mu)}{2}
 \nonumber
 \;,
\end{eqnarray}
where the likelihood in each bin is
\begin{equation}
 \label{equ:likelihood_ngc1068_profiled}
 \mathcal{L}_{ij}
 (\boldsymbol\theta_{\rm ast},
 \boldsymbol{\theta}_{\rm{mix}},
 \boldsymbol\theta_{\rm dec}, 
 \mathcal{N}^\mu,
 T)
 =
 \frac{\mu_{ij,\rm t}^{n_{ij,\rm t}}}{n_{ij,\rm t}!}
 e^{-\mu_{ij,\rm t}} \;, 
\end{equation}
and, on the right-hand side, $\mu_{ij} \equiv \mu_{ij}(\boldsymbol\theta_{\rm ast}, \boldsymbol{\theta}_{\rm{mix}},  \boldsymbol\theta_{\rm dec}, \allowbreak \mathcal{N}^\mu, T)$ is computed using \equ{event_rate_ngc1068}.  In \equ{likelihood_ngc1068}, $\chi^2(\boldsymbol{\theta}_{\rm{mix}})/2$ and $\chi^2(\mathcal{N}^\mu)/2$ are, respectively, pull terms that represent our prior knowledge of the mixing and atmospheric-background parameters.  In our baseline scenario, we do not include pull terms for the astrophysical parameters to avoid introducing bias in our results on the decay parameters.  But, in our alternative scenario with a tight prior on the flux normalization, we add a pull term on it, \ie, $\chi^2(\Phi_0)/2$; see Sec.~\ref{sec:bounds_ngc1068-flux} and Table~\ref{tab:ngc_1068}.  For the mixing parameters, we use as priors their distributions from the recent global fit to oscillation data, NuFIT~5.2~\cite{Esteban:2020cvm,NuFit5.2}, assuming normal neutrino mass ordering.  For the atmospheric background normalization, we use a flat prior centered on the best-fit value reported in the IceCube analysis~\cite{IceCube:2022der}, and with a width of two decades below and above this value.  We tested that using reasonably wider or narrower priors does not affect our results appreciably, as the normalization of the background is constrained well by measurements in sky patches far from NGC 1068.

Since we are only interested in the sensitivity to the decay parameters, we treat the astrophysical, mixing, and atmospheric parameters as nuisance, and profile the likelihood function over them, \ie,
\begin{equation}
 \tilde{\mathcal{L}}
 (\boldsymbol{\theta}_{\rm dec},
 T)
 =
 \mathrm{max}_{\boldsymbol{\theta}_{\rm ast},\boldsymbol{\theta}_{\rm mix},\mathcal{N}^\mu}
 \mathcal{L}
 (\boldsymbol\theta_{\rm ast},
 \boldsymbol{\theta}_{\rm{mix}},
 \boldsymbol\theta_{\rm dec}, 
 \mathcal{N}^\mu,
 T) \;.
\end{equation}
We find that the introduction of neutrino decay does not improve the fit to the data.  The best-fit value of the decay parameters, $\hat{\boldsymbol{\theta}}_{\rm dec}=\left\{0,0\right\}$, corresponds to no decay of $\nu_2$ and $\nu_3$. Therefore, we compute lower bounds on the lifetimes $\tau_2/m_2$ and $\tau_3/m_3$, \ie, upper bounds on $\boldsymbol{\theta}_{\rm dec}$. We do this via the test statistic
\begin{equation}
 \label{equ:test_statistic_ngc1068}
 \Lambda(\boldsymbol{\theta}_\mathrm{dec}, T)
 =
 2
 \left[
 \log\tilde{\mathcal{L}}(\hat{\boldsymbol{\theta}}_{\rm dec}, T)
 -
 \log\tilde{\mathcal{L}}({\boldsymbol{\theta}_{\rm dec}}, T)
 \right] \;.
\end{equation}
By virtue of Wilks' theorem~\cite{Wilks:1938dza}, under the hypothesis that the true decay parameters are $\boldsymbol{\theta}_{\rm dec}$, the test statistic is expected to be distributed as a chi-squared variable with two degrees of freedom, corresponding to the two decay parameters, $(\tau_2/m_2)^{-1}$ and $(\tau_3/m_3)^{-1}$.

\smallskip

\textbf{\textit{Using projected data.---}}The fundamental limitation in probing the decay of neutrinos from NGC 1068 is that, because they are so far detected only as tracks, we are limited to searching for the relatively small jump in flux normalization that decay could induce (Sec.~\ref{sec:nu_prop_with_decay-energy_spectrum}).  This is aggravated by the fact that---at least in our baseline scenario---there is no prior information as to what the flux normalization is before it was possibly affected by decay.  Barring gaining access to the flavor composition of the flux from NGC 1068---which we do not consider here (see, however, \Refe~\cite{Valera:2023bud})---the only way to improve the precision with which decay can be probed in the future is to measure the energy spectrum more precisely. 

We explore this scenario by increasing the number of detected events (we mention other possible improvements in Sec.~\ref{sec:discussion}).  We repeat our analysis above, but now using the cumulative sample of tracks detected, up to the year 2035, collectively by IceCube and upcoming neutrino telescopes Baikal-GVD, IceCube-Gen2, KM3NeT, and P-ONE, each computed within the PLE$\nu$M framework~\cite{Schumacher:2021hhm} (Sec.~\ref{sec:bounds_ngc1068-event_rates}).  We scale and rotate the present-day IceCube sensitivity to a neutrino point source in order to simulate the sensitivity of the other telescopes, which have sizes and geographical locations different from IceCube.  The total number of detected tracks is the sum of their contributions. 

In our projections, we assume for the true $\nu_\mu + \bar{\nu}_\mu$ flux the best-fit power-law flux reported by IceCube~\cite{IceCube:2022der}, $\Phi_{\nu_\mu + \bar{\nu}_\mu} = \Phi_0 [E_\nu/(1~{\rm TeV})]^{-\gamma}$, with $\Phi_0 = 5 \cdot 10^{-14}$~GeV$^{-1}$~cm$^{-2}$~s$^{-1}$ and $\gamma = 3.2$, and a pion-beam flavor composition, which allows us to compute the all-flavor flux using the present-day best-fit values of the neutrino mixing parameters.  We use Asimov data sets, \ie, we assume that the future observed event rates coincide with the expected rates ~\cite{Cowan:2010js}.  The rest of the analysis proceeds as when using present-day data.  


\subsection{Results}
\label{sec:bounds_ngc1068-results}

\begin{figure}[t!]
 \includegraphics[width=\columnwidth]{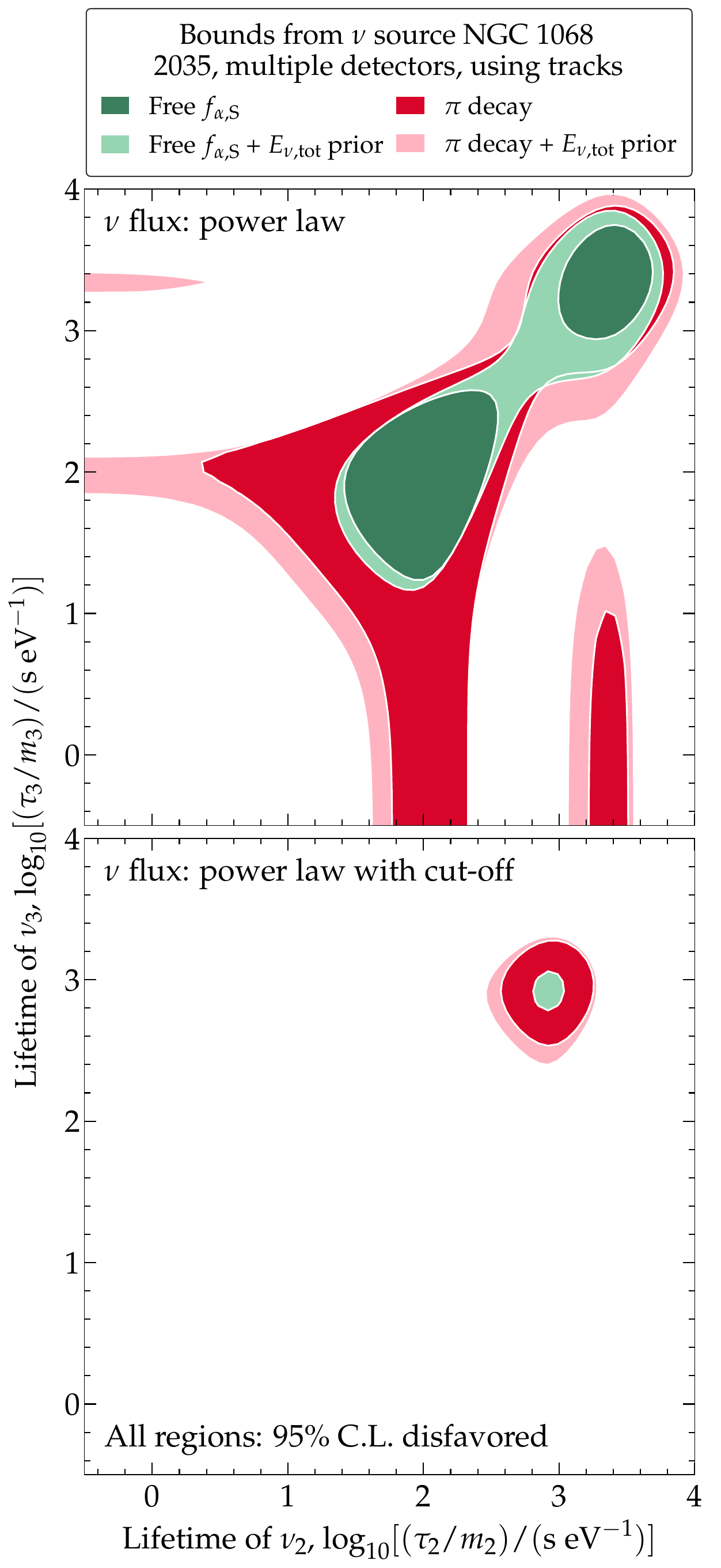}
 \caption{\textbf{\textit{Bounds on the neutrino lifetime using high-energy neutrinos from NGC 1068.}}  All bounds are forecast to the year 2035, from tracks detected by multiple upcoming neutrino telescopes; no bounds can be placed using present IceCube data.  
 Bounds are profiled over all other model parameters, allowing for free flavor composition at the sources (``free $f_{\alpha, {\rm S}}$''), fixing it to pion decay, and possibly constraining the total neutrino energy (``$E_{\nu, {\rm tot}}$ prior'').  
 {\it Top:} Assuming a power-law neutrino energy spectrum. {\it Bottom:} Assuming a power law with a high-energy cut-off.  See Sec.~\ref{sec:bounds_ngc1068} for details, \Refe~\cite{github_repo} for plot data.
 \textit{Future neutrino lifetime bounds from NGC 1068 hinge on astrophysical unknowns.}
 \vspace*{-1cm}
 }  
 \label{fig:bounds_ngc1068}
\end{figure}

Figure~\ref{fig:bounds_ngc1068} shows our resulting two-dimensional joint bounds on the lifetimes of $\nu_2$ and $\nu_3$, for the scenarios listed in Table~\ref{tab:ngc_1068}.  This table, and \figu{1d_bounds_summary}, show also the one-dimensional bounds on the individual lifetimes that, however, discard important correlations between $\tau_2/m2$ and $\tau_3/m_3$, as we explain below.  

\textbf{\textit{Using present-day IceCube observations of NGC 1068 it is not possible to constrain neutrino decay.}}  All of the results shown in \figu{bounds_ngc1068} are projections, not present-day results.  Today, the precision with which the NGC 1068 neutrino spectrum is measured is too poor to spot the step-like factor-of-two-thirds reduction, at most, in the flux normalization that neutrino decay could induce (\figu{flux_ngc1068} and Sec.~\ref{sec:nu_prop_with_decay-diffuse_vs_point}).  This is true in all of our eight analysis scenarios, even in the most optimistic one where we assume a power-law spectrum, fix the flavor composition to pion-beam, and impose a prior on the total neutrino energy emitted.  

The above limitation is representative not just of searches for neutrino decay using neutrinos from NGC 1068, but broadly of studies of neutrino physics and astrophysics that rely predominantly on spotting small changes to the normalization of the neutrino flux.  Underlying this limitation are the astrophysical unknowns and the insufficient precision of present-day measurements of the neutrino flux.  Ignoring either of these aspects would result in unrealistically high sensitivity.

Figure~\ref{fig:bounds_ngc1068} shows that, in the future, it should be possible to constrain the $\nu_2$ and $\nu_3$ lifetimes, at least under some analysis choices (more on this below).  Since our analysis relies on identifying a change in the flux normalization within the energy window in which IceCube observes NGC 1068, our results disfavor lifetimes in the range $L / E_\nu \approx 10^2$--$10^3$~s~eV$^{-1}$, where $L = 14$~Mpc and $E_\nu = 1$--10 TeV.  Neutrino lifetimes below this range would change the flux normalization at energies above the IceCube energy window; lifetimes above this range, at energies below the IceCube energy window.  In either case, the transition becomes undetectable and so there is no sensitivity to lifetimes outside the above range.

\begin{figure}[t!]
 \includegraphics[width=\columnwidth]{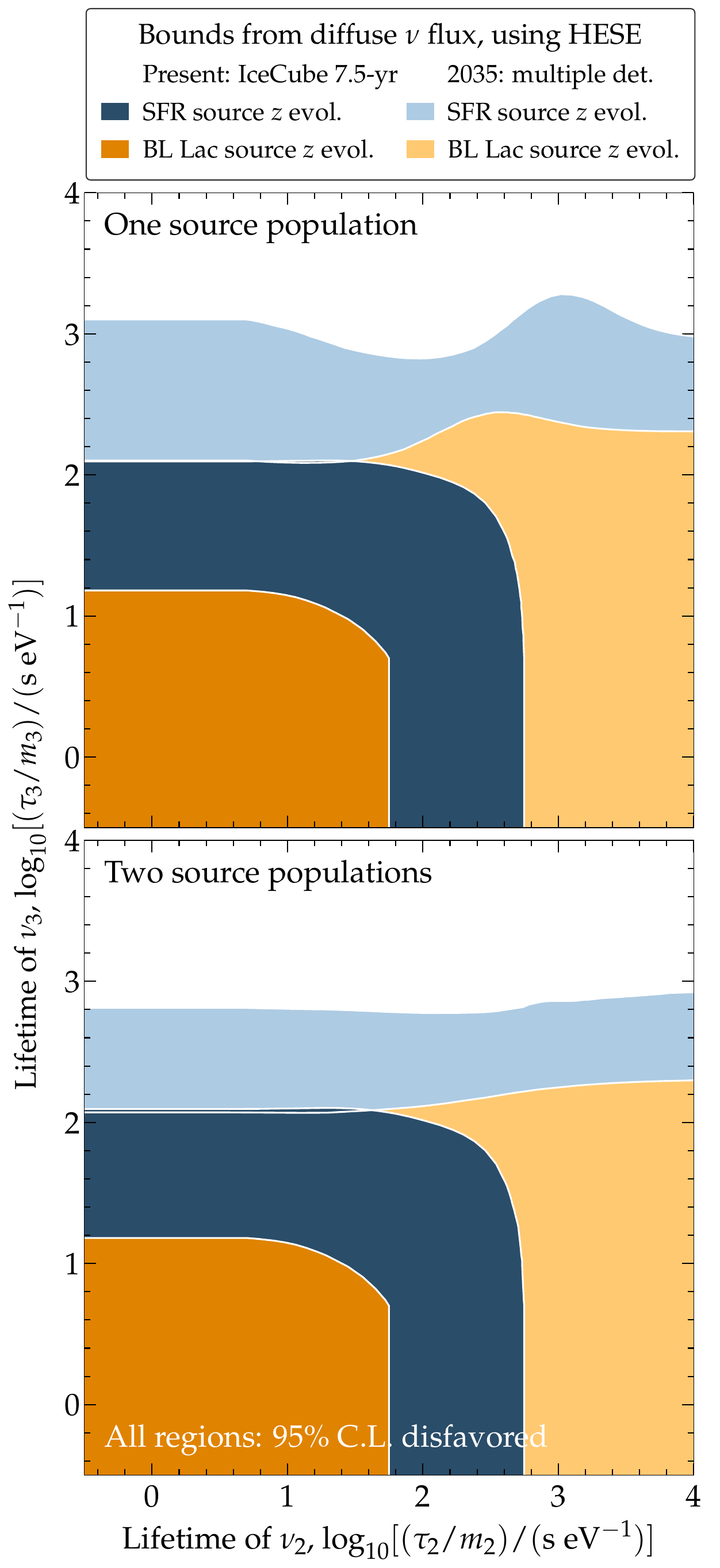}
 \caption{\textbf{\textit{Bounds on the neutrino lifetime using the diffuse flux of high-energy astrophysical neutrinos.}}  Bounds are from HESE events.  Present bounds are from the IceCube 7.5-year sample~\cite{IceCube:2020wum, IC75yrHESEPublicDataRelease}; bounds in the year 2035 are from events detected by multiple upcoming neutrino telescopes.  Bounds are profiled over all other model parameters, for two possible choices of the redshift evolution of the source number density: star-formation rate (SFR) and BL Lac.  {\it Top:} Assuming one source population.  {\it Bottom:} Assuming two populations.  See Sec.~\ref{sec:bounds_diffuse} for details, \Refe~\cite{github_repo} for plot data.  \textit{Lifetime bounds from the diffuse flux are  sensitive to the redshift distribution of neutrino sources.}
 \vspace*{-1cm}
 }
 \label{fig:bounds_diffuse}
\end{figure}

Indeed, \figu{bounds_ngc1068} shows that, assuming the PL spectrum, if the flavor composition is allowed to float freely and no prior on $E_{\nu, {\rm tot}}$ is adopted, two disconnected regions are excluded, with $\tau_2/m_2 \approx \tau_3/m_3$ either above $10^3$~s~eV$^{-1}$, or close to $10^2$~s~eV$^{-1}$.  
\textbf{\textit{However, the projected constraints on neutrino lifetimes based on the neutrinos from NGC 1068 are highly dependent on our analysis choices for the shape of the neutrino spectrum, flavor composition, and the use of a prior on the total energy emitted in neutrinos.}}  

Naturally, fixing the flavor composition to be pion-beam \textit{vs.}~letting it float freely allows us to spot smaller changes to the flux normalization and, therefore, to disfavor larger regions of lifetime.  Figure~\ref{fig:bounds_ngc1068} shows that, assuming a PL spectrum, our analysis disfavors lifetimes of  $100$~s~eV$^{-1}$~$\lesssim \tau_2/m_2 \approx \tau_3/m_3 \lesssim 5000$~s~eV$^{-1}$, where all neutrinos decay below a transition energy in the $1$--$10$~TeV range.  In addition, assuming a known flavor composition allows us to exclude the vertical and horizontal regions in \figu{bounds_ngc1068} corresponding to $\tau_2/m_2 \simeq 100$~s~eV$^{-1}$ or $1000$~s~eV$^{-1}$ while $\tau_3/m_3 \ll 100$~s~eV$^{-1}$, and vice versa, corresponding to the cases where either $\nu_2$ or $\nu_3$ decay completely and the decay of the remaining species induces a flux transition within the IceCube energy window.  Similarly, adopting a prior on $E_{\nu, {\rm tot}}$ enlarges the region of lifetimes disfavored, since it also allows us to spot smaller changes to the flux normalization.  

Figure~\ref{fig:bounds_ngc1068} shows that the largest effect on the lifetime bounds is due to switching between using the PL model \textit{vs.}~the PC models of the neutrino energy spectrum.  Our bounds nearly disappear when using the latter model because the high-energy cut-off in the spectrum may mimic the step-like transition in the flux normalization that would be induced by neutrino decay, for some choices of the value of the cut-off energy, $E_{\nu, {\rm cut}}$.  Once more, this reflects the critical importance of accounting for astrophysical model uncertainties in searches for new physics.

In our most conservative scenario---adopting the PL model, freely floating flavor composition, and no prior on $E_{\nu, {\rm tot}}$---no lifetime bounds can be drawn.  While egregious, this scenario is likely too extreme, and in reality we can likely do better.  By fixing instead the flavor composition to pion-beam---as motivated by models of NGC 1068 (Sec.~\ref{sec:bounds_ngc1068-flux})---we can exclude $\tau_2/m_2 \approx \tau_3/m_3 \approx 1000$~s~eV$^{-1}$.  This corresponds to a jump in the flux normalization at around $1$~TeV, where it is easier to spot due to the neutrino flux being higher there.  While the disfavored region is small, \figu{bounds_overview} shows that it is unreachable by searches for neutrino decay based on the diffuse neutrino flux (Sec.~\ref{sec:bounds_diffuse}), where neutrino energies are typically higher and so are sensitive to shorter lifetimes.


\subsection{Future improvements}
\label{sec:bounds_ngc1068-future}

As pointed out earlier, the main limitation to testing neutrino decay using NGC 1068 is that, because today we only observe it via tracks, we are limited to looking for decay-induced features on the neutrino energy spectrum.  Combining the detection of tracks---made mainly from $\nu_\mu$---and cascades---made by all flavors (Sec.~\ref{sec:bounds_diffuse-hese})---from NGC 1068 could circumvent the above limitation by providing an additional probe of neutrino via the flavor composition (Sec.~\ref{sec:nu_prop_with_decay-flavor_comp}).  However, because the pointing resolution of IceCube cascades is significantly worse than that of tracks---of tens of degrees \textit{vs.}~sub-degree---including them in our analysis, today, would come at the cost of using a larger atmospheric neutrino background. 

However, the advent of next-generation water-Cherenkov detectors like Baikal-GVD, KM3NeT, P-ONE, and TRIDENT opens up the possibility of observing astrophysical point neutrino sources using also cascades.  There, cascades are expected to have a better pointing resolution than in IceCube thanks to the photon scattering length being longer in water than in ice.  This would grant us access to tests of neutrino decay using NGC 1068 not only via its neutrino energy spectrum but also via its flavor composition.  Reference~\cite{Valera:2023bud} showed promising preliminary results for this possibility, but an in-depth analysis, possibly using improved flavor-measurement techniques~\cite{Li:2016kra, Steuer:2017tca, Farrag:2023jut}, is beyond the scope of the present work.


\section{Neutrino lifetime bounds from the diffuse neutrino flux}
\label{sec:bounds_diffuse}

The observation of neutrinos from NGC 1068 cannot at present exclude neutrino decay, but may do so in the future (Sec.~\ref{sec:bounds_ngc1068}).  In contrast, the observation of the diffuse flux of high-energy astrophysical neutrinos in IceCube can probe neutrino decay already today~\cite{Pakvasa:1981ci, Beacom:2002vi, Barenboim:2003jm, Beacom:2003nh, Beacom:2003zg, Meloni:2006gv, Maltoni:2008jr, Bustamante:2010nq, Baerwald:2012kc, Pakvasa:2012db, Dorame:2013lka, Pagliaroli:2015rca, Bustamante:2015waa, Huang:2015flc, Shoemaker:2015qul, Bustamante:2016ciw}.  

The main competitive advantage of using the diffuse flux is that it gives us access not only to the neutrino energy spectrum---like tracks do for NGC 1068---but also to the flavor composition, both of which may be affected by neutrino decay (Sec~\ref{sec:nu_prop_with_decay-diffuse_vs_point}).  

However, because the identity and number of neutrino sources, and the neutrino production mechanism, are unknown, our claims regarding neutrino decay are population-averaged.  Below, we explore several representative possibilities spanning the breadth of astrophysical unknowns and show that their influence on our results, while appreciable, is not decisive.

We describe the procedure that we follow to search IceCube data for evidence of neutrino decay, and the constraints on neutrino lifetime that we place.  First, we do this for the present-day public 7.5-year public sample of IceCube High-Energy Starting Events (HESE)~\cite{IceCube:2020wum}.  Then, we show forecasts based on  projected combined measurements by multiple upcoming neutrino telescopes.


\subsection{Diffuse high-energy astrophysical neutrino flux}
\label{sec:bounds_diffuse-flux}

Despite the discovery of the first few sources of high-energy astrophysical neutrinos~\cite{IceCube:2018cha, Stein:2020xhk, Reusch:2021ztx, IceCube:2022der, IceCube:2023ame}, including NGC 1068, the origin of the bulk of them is unknown.  Because the diffuse flux is largely isotropic, the astrophysical sources responsible for it are likely largely extragalactic.  They are purportedly cosmic accelerators capable of boosting cosmic-ray protons and nuclei up to energies of at least tens of PeV, and possibly higher~\cite{AlvesBatista:2019tlv}.  

Today, several candidate source classes are under consideration, including active galactic nuclei (AGN), gamma-ray bursts (GRBs), starburst galaxies (SBGs), and galaxy clusters (GCs).  Competing theoretical predictions of the neutrino flux from them differ in the specifics, but share a dependence on key factors like the acceleration mechanism, the properties of the ambient matter and radiation with which cosmic rays interact to make neutrinos, and the intensity and configuration of the magnetic fields harbored by the sources.  Presently, there is little to no direct experimental information about these properties.  Rather than exploring the detailed predictions from competing models, we adopt a phenomenological prescription that captures common features of their energy spectrum and flavor composition.

For the diffuse neutrino spectrum, we adopt a power law, $\Phi_{\nu_\alpha} \propto E_\nu^{-\gamma}$, where $\gamma > 2$; this shape is commonly used in the literature; see, \eg, \Refes~\cite{Winter:2012xq, Fiorillo:2022rft, Fiorillo:2024jqz} and references therein.  Using the recent 7.5-year HESE sample---the same one we use (more on this later)---the IceCube Collaboration reported~\cite{IceCube:2020wum} $\gamma = 2.87_{-0.19}^{+0.20}$; see \figu{flux_diffuse}.  [Other analyses based on different event samples, and covering different but overlapping energy ranges, find harder, but compatible spectra; \ie, through-going muons~\cite{IceCube:2021uhz} yield $\gamma = 2.37 \pm 0.09$, starting tracks~\cite{Abbasi:2024jro} yield $\gamma = 2.58_{-0.09}^{+0.10}$, and a combined analysis using different types of event samples~\cite{IceCube:2015gsk}, now somewhat outdated, yields $\gamma = 2.50 \pm 0.09$ (\Refe~\cite{Naab:2023xcz} has a preliminary update, with comparable results)].  Later, in our statistical analysis based on present-day data, we let the value of $\gamma$ float and be determined by the comparison to real data.  When making forecasts, we generate mock event samples by adopting for the true value of $\gamma$ its best-fit value from the 7.5~year HESE analysis~\cite{IceCube:2020wum}, and then let its value be determined by the comparison to mock data.   

A power-law spectrum is particularly well-motivated in source candidate classes where we expect neutrino production to occur primarily via $pp$ interactions, \eg, SBGs and GCs.  In sources where we expect it to occur primarily via $p\gamma$ interactions, \eg, AGN and GRBs, the neutrino spectrum should exhibit a ``bump'' around a characteristic energy set by the shape of the target photon spectrum; see Fig.~2 in \Refe~\cite{Fiorillo:2022rft} for an overview.  Today, arguably the most convincing argument in favor of choosing to model the neutrino spectrum as a power law {\it vs.}~choosing alternative, more complex shapes is its consistency with the IceCube observations.  Given present-day measurement uncertainties, these alternatives are so far not significantly preferred; see, \eg, \Refes~\cite{IceCube:2015gsk, IceCube:2020wum, IceCube:2021uhz, Fiorillo:2022rft, Abbasi:2024jro}.

Thus, we compute the diffuse neutrino flux by considering a population of identical, nondescript astrophysical sources, each one emitting $\nu_\alpha + \bar{\nu}_\alpha$ with a spectrum
\begin{equation}
 \frac{dN_\alpha}{dE_\nu dt}
 \propto
 f_{\alpha, {\rm S}} 
 E_\nu^{-\gamma} \;.
\end{equation}
We assume that the number density of sources, $\rho_{\rm src}$, evolves with redshift, and compute the diffuse flux at Earth including the effects of neutrino decay, $\Phi_{\nu_\alpha}$ in \equ{diffuse_integration}.  We treat its all-flavor normalization constant at 100~TeV, $\Phi_0$, as a free parameter whose value is determined by observations, real or projected.  For the flavor composition at the sources, $(f_e, f_\mu, f_\tau)_{\rm S}$, we assume there is no production of $\nu_\tau$, \ie, $f_{\tau, {\rm S}} = 0$,  like we did for NGC 1068.  For our baseline scenario, we consider a single population (1P) of neutrino sources responsible for the diffuse flux, and compute it using \equ{diffuse_integration}. 

The diffuse flux that we compute is isotropic, but, in reality, about 10\% of the observed diffuse flux comes from the Galactic Plane~\cite{IceCube:2023ame, Bustamante:2023iyn} at tens of TeV.  Because neutrinos from the Galactic Plane travel a much shorter path to Earth, the effect of decay on them is weaker than on neutrinos of extragalactic origin, and should be treated separately, or their incoming directions should be masked out.  However, in our analysis we assume that all neutrinos are extragalactic, since there is presently no publicly available detailed directional information on the neutrinos detected from the Galactic Plane that we can incorporate in our analysis.

For the redshift evolution of the source number density, we adopt as our baseline assumption the star-formation rate (SFR)~\cite{Hopkins:2006bw}.  This places most of the sources at redshifts $z \sim 1$, corresponding to a few Gpc, thus lengthening the time of neutrino propagation to Earth and enhancing the potential impact of neutrino decay during it.  As a more conservative choice, we also perform an analyses where we adopt a source evolution like that of BL Lac objects, a type of AGN, as extracted from Fig.~5 of \Refe~\cite{Capanema:2020oet} (in turn obtained from \Refe~\cite{Ajello:2013lka}).  With it, the source density rapidly falls with redshift, placing more sources closer to Earth.

In addition to the 1P scenario above, we explore the alternative scenario of two contributing populations of neutrino sources (2P), each emitting neutrinos with a different spectrum and flavor composition.  We examine whether their interplay leads to features in their combined energy spectrum and flavor composition that could cloud the presence of neutrino decay.  The neutrino flux is the superposition of the power-law fluxes from two populations with independent normalization constants $\Phi_{0, 1}$ and $\Phi_{0, 2}$, spectral indices $\gamma_1$ and $\gamma_2$, and flavor compositions $f_{\alpha, 1}$ and $f_{\alpha, 2}$.  For simplicity, we use a common source density evolution for both populations.  

Table~\ref{tab:diffuse} summarizes the eight scenarios that we explore using the diffuse flux for the sensitivity to neutrino decay: two astrophysical production models (1P and 2P), two choices of the source redshift evolution (SFR and BL Lac), and two choices for the treatment of the flavor composition at the sources (freely floating and fixed to pion-beam).  For each scenario, we compute constraints on $\tau_2/m_2$ and $\tau_3/m_3$.

\begingroup
\squeezetable
\begin{table*}[t!]
 \begin{ruledtabular} 
  \caption{\label{tab:diffuse}\textbf{\textit{Constraints on the neutrino lifetime from the diffuse flux of high-energy astrophysical neutrinos.}}  The constraints are on the lifetimes of $\nu_2$ and $\nu_3$, computed using HESE events, using present-day IceCube data and the projected, cumulative detection by the combination of neutrino telescopes Baikal-GVD, IceCube, IceCube-Gen2, KM3NeT, P-ONE, TAMBO, and TRIDENT by the year 2035.  The constraints are shown after profiling over all other model parameters, for our eight scenarios of the number of neutrino source populations (1P and 2P), the redshift evolution of their number density (SFR and BL Lac), and the treatment of the flux composition at the sources, $f_{\rm S}$ (freely floating and fixed to the pion-beam composition).  Two-dimensional constraints are preferred because they depict the correlation between $\tau_2/m_2$ and $\tau_3/m_3$, which is not captured by the one-dimensional constraints.  Blank entries, marked with ``$\cdots$'', represent scenarios where no constraint can be placed.  See \figu{1d_bounds_summary} for a graphical comparison of all one-dimensional constraints and Secs.~\ref{sec:nu_prop_with_decay} and \ref{sec:bounds_diffuse} for details.}
  \centering
  \renewcommand{\arraystretch}{1.7}
  \begin{tabular}{cccccccccccccccc}
   \multirow{5}{*}{\makecell{Number of \\ neutrino \\ source \\ populations}}        &
   \multirow{5}{*}{\makecell{Source \\ density \\ redshift \\ evolution, \\ $\rho_{\rm src}$}}            &
   \multicolumn{14}{c}{Neutrino lifetime disfavored at 95\%~C.L., $\tau_i/m_i$ [s~eV$^{-1}$]} \\
   \cline{3-16}
   &
   &
   \multicolumn{7}{c}{Present (7.5-yr IceCube HESE sample)}         &
   \multicolumn{7}{c}{Projections (year 2035, multiple detectors, HESE)} \\
   \cline{3-9}
   \cline{10-16}
   &
   &
   \multicolumn{4}{c}{$f_{\rm S}$ free}                       &
   \multicolumn{3}{c}{$f_{\rm S}$ fixed to $\pi$ decay}       &
   \multicolumn{4}{c}{$f_{\rm S}$ free}                       &
   \multicolumn{3}{c}{$f_{\rm S}$ fixed to $\pi$ decay}       \\
   \cline{3-6}
   \cline{7-9}
   \cline{10-13}
   \cline{14-16}
   &
   &
   \multirow{2}{*}{\makecell{$\frac{\tau_3}{m_3}$~{\it vs.}~$\frac{\tau_2}{m_2}$ \\[0.5em] (preferred)}}   &   
   \multirow{2}{*}{$\frac{\tau_2}{m_2}$\footnote{\label{fnote5}One-dimensional lower limit after profiling over $\tau_3/m_3$.  See also \figu{1d_bounds_summary}.}}    &
   \multirow{2}{*}{$\frac{\tau_3}{m_3}$\footnote{\label{fnote6}One-dimensional lower limit after profiling over $\tau_2/m_2$.  See also \figu{1d_bounds_summary}.}}    &
   \multirow{2}{*}{$\frac{\tau_2}{m_2} = \frac{\tau_3}{m_3}$\footnote{\label{fnote7}One-dimensional lower limit obtained assuming $\tau_2/m_2 = \tau_3/m_3$.  See also \figu{1d_bounds_summary}.}}                                                 &   
   \multirow{2}{*}{$\frac{\tau_2}{m_2}$}                                       &
   \multirow{2}{*}{$\frac{\tau_3}{m_3}$}                                       &
   \multirow{2}{*}{$\frac{\tau_2}{m_2} = \frac{\tau_3}{m_3}$}                  &   
   \multirow{2}{*}{\makecell{$\frac{\tau_3}{m_3}$~{\it vs.}~$\frac{\tau_2}{m_2}$ \\[0.5em] (preferred)}}   &   
   \multirow{2}{*}{$\frac{\tau_2}{m_2}$}                                       &
   \multirow{2}{*}{$\frac{\tau_3}{m_3}$}                                       &
   \multirow{2}{*}{$\frac{\tau_2}{m_2} = \frac{\tau_3}{m_3}$}                  &   
   \multirow{2}{*}{$\frac{\tau_2}{m_2}$}                                       &
   \multirow{2}{*}{$\frac{\tau_3}{m_3}$}                                       &
   \multirow{2}{*}{$\frac{\tau_2}{m_2} = \frac{\tau_3}{m_3}$}                  \\
   \\
   \hline
   1P                                                         &
   SFR                                                        &
   \figu{bounds_diffuse}                                      &
   $\cdots$                                                   &
   $\cdots$                                                   &
   $\leq$~219                                                 &
   $\cdots$                                                   &
   $\leq$~126                                                 &
   $\leq$~550                                                 &
   \figu{bounds_diffuse}                                      &
   $\cdots$                                                   &
   $\leq$~1381                                                &
   $\leq$~2631                                                &
   $\cdots$                                                   &
   $\leq$~1260                                                &
   $\leq$~1996                                                \\
   1P                                                         &
   BL Lac                                                     &
   \figu{bounds_diffuse}                                      &
   $\cdots$                                                   &
   $\cdots$                                                   &
   $\leq$~29                                                  &
   $\cdots$                                                   &
   $\leq$~20                                                  &
   $\leq$~105                                                 &
   \figu{bounds_diffuse}                                      &
   $\cdots$                                                   &
   $\leq$~316                                                 &
   $\leq$~550                                                 &
   $\cdots$                                                   &
   $\leq$~316                                                 &
   $\leq$~501                                                 \\
   2P                                                         &
   SFR                                                        &
   \figu{bounds_diffuse}                                      &
   $\cdots$                                                   &
   $\cdots$                                                   &
   $\leq$~219                                                 &
   $\cdots$                                                   &
   $\leq$~126                                                 &
   $\leq$~550                                                 &
   \figu{bounds_diffuse}                                      &
   $\cdots$                                                   &
   $\leq$~1259                                                &
   $\leq$~2400                                                &
   $\cdots$                                                   &
   $\leq$~1260                                                &
   $\leq$~1996                                                \\
   2P                                                         &
   BL Lac                                                     &
   \figu{bounds_diffuse}                                      &
   $\cdots$                                                   &
   $\cdots$                                                   &
   $\leq$~26                                                  &
   $\cdots$                                                   &
   $\leq$~20                                                  &
   $\leq$~105                                                 &
   \figu{bounds_diffuse}                                      &
   $\cdots$                                                   &
   $\leq$~316                                                 &
   $\leq$~182                                                 &
   $\cdots$                                                   &
   $\leq$~316                                                 &
   $\leq$~417                                                 \\  
  \end{tabular}
 \end{ruledtabular}
\end{table*}
\endgroup


\subsection{High-Energy Starting Events (HESE)}
\label{sec:bounds_diffuse-hese}

The measurement of the diffuse flux of high-energy astrophysical neutrinos in IceCube adopts different strategies to mitigate the background of high-energy atmospheric neutrinos.  We focus on HESE events~\cite{IceCube:2013low}, \ie, events where the neutrino-nucleon interaction occurs inside the fiducial volume of the detector and the outer layer of photomultiplier strings in the detector is used as a veto to reject the coincident detection of an accompanying muon that would tag an event as being of atmospheric, rather than astrophysical origin~\cite{Schonert:2008is, Gaisser:2014bja, Arguelles:2018awr}.  As a result, the samples of HESE events are among the most astrophysically pure ones available, and are therefore well-suited for our purposes. 

Compared to the through-going and starting tracks that we used for NGC 1068, HESE events are rare. The public 7.5-year IceCube HESE sample~\cite{IceCube:2020wum, IC75yrHESEPublicDataRelease} that we use contains 102 detected events between tens of TeV and a few PeV.  [A more recent sample~\cite{IceCube:2023sov, IC12yrHESEPublicDataRelease}, using 12 years, contains 164 events.  However, because it has no publicly available accompanying sample of Monte Carlo events that we can use in our statistical analysis (more on this later), we use the 7.5-year sample instead.]

At these energies, neutrinos typically interact in the detector via neutrino-nucleon ($\nu N$) deep inelastic scattering.  In it, the interacting neutrino scatters off of one of the quarks or gluons of a proton or neutron in the ice, which is broken up into final-state hadrons, $X$, as a result.  Interactions can be either neutral-current, when mediated by a $Z$ boson, \ie, $\nu_\alpha + N \to \nu_\alpha + X$, or charged-current, when mediated by a $W$ boson, \ie, $\nu_\alpha + N \to l_\alpha + X$.  In both cases, the final-state charged particles radiate Cherenkov light that, in IceCube, is picked up by photomultipliers buried in the Antarctic ice.  From the amount of light detected, and from its spatial and temporal profiles, it is possible to infer the energy, arrival direction, and, to an extent, the flavor of the interacting neutrino (more on this below).  

A HESE event is classified in one of three topologies---cascades, tracks, and double cascades---according to  its light profile.  Cascades (c) are made primarily by the charged-current interaction of $\nu_e$ and $\nu_\tau$, where the particle showers triggered by the final-stated hadrons and the final-state electron or tau are superimposed and detected as one.  They are also made by the neutral-current interaction of neutrinos of all flavors, although the neutral-current cross section is smaller than the charged-current one.  Tracks (t) are made primarily by the charged-current interaction of $\nu_\mu$, where the final-state muon leaves a km-scale track of Cherenkov light in its wake that is easily identifiable.  They are also made by the charged-current interaction of $\nu_\tau$, which generate a tau that decays into a muon about 17\% of the time~\cite{ParticleDataGroup:2022pth}.  Double cascades (dc)~\cite{Learned:1994wg} are made by the charged-current interaction of $\nu_\tau$ where a first cascade is produced by the $\nu_\tau N$ interaction and a second one, by the hadronic decay of the final-state tau. The majority of detected events are cascades; double cascades are the rarest. 

The classification above reveals why it is not possible to firmly infer the flavor of the neutrino in a specific HESE event: a cascade could have been made by a neutrino of any flavor, and a track could have been made by a $\nu_\mu$ or a $\nu_\tau$.  However, as proposed first by \Refe~\cite{Beacom:2003nh}, from the relative number of detected events of each topology, IceCube analyses~\cite{IceCube:2015rro, IceCube:2015gsk, IceCube:2018pgc, IceCube:2020fpi} (see also, \eg, \Refes~\cite{Mena:2014sja, Palomares-Ruiz:2015mka, Palladino:2015vna, Vincent:2016nut}) infer the flavor composition of the neutrino flux, statistically.  Thus,  HESE events---unlike the through-going~\cite{IceCube:2021uhz} or starting tracks~\cite{Abbasi:2024jro} from NGC 1068---allow us to look for signs of neutrino decay jointly in the neutrino energy spectrum and flavor composition; see, \eg, \Refes~\cite{Beacom:2002vi, Bustamante:2015waa, Shoemaker:2015qul, Bustamante:2016ciw}.  We comment later on combining measurements of the diffuse flux via HESE events and through-going tracks (Sec.~\ref{sec:bounds_diffuse-future}).


\subsection{Generating HESE event samples}
\label{sec:bounds_diffuse-hese_mc}

To accurately model the capability of IceCube to detect high-energy astrophysical neutrinos via HESE events, including the above nuances, we use the Monte Carlo (MC) data sample~\cite{IC75yrHESEPublicDataRelease} that is provided by the IceCube Collaboration together with the 7.5-year HESE data release~\cite{IceCube:2020wum}. The  MC sample contains a large number of simulated HESE events that were generated using the same detailed detector response used in the analysis of the 7.5-year  sample by the IceCube Collaboration.  We use it to compute mock samples of detected events for different choices of the neutrino flux, including ones affected by neutrino decay, which we later contrast against the present or projected samples.

The events in the MC sample were generated assuming a reference diffuse high-energy astrophysical neutrino flux~\cite{IC75yrHESEPublicDataRelease} of $\Phi_{\alpha, \rm{ref}} = (\Phi_{0, {\rm ref}} / 3) (E_\nu/100~{\rm TeV})^{-2.87}$ for each of three flavors, with $\Phi_{0, {\rm ref}} = 5.68 \cdot 10^{-18}$ GeV$^{-1}$~cm$^{-2}$~s$^{-1}$~sr$^{-1}$, and for an exposure time $T_{\rm ref} = 2635$~days.  The $k$-th MC event in the sample consists of the flavor, energy, and direction of the simulated neutrino, $\alpha^k$, $E_\nu^k$ and $\cos\theta_{z}^k$, respectively, where $\theta_z$ is measured with respect to the zenith, and of the event topology (cascade, track, double cascade), reconstructed deposited energy and direction of the ensuing HESE event, t$^k$, $E_\mathrm{dep}^k$ and $\cos\theta_{z,\mathrm{rec}}^k$, respectively.  In addition, the event has an associated reference MC weight $w_\mathrm{ref}^k$. 

To generate a mock sample of HESE events due to an arbitrary diffuse neutrino flux, $\Phi_\alpha$, such as the one computed in \equ{diffuse_integration}, and for arbitrary exposure time, $T$, we reweigh the MC events by defining updated weights according to their neutrino energy and flavor, \ie,
\begin{equation}
 w^k(E_\nu^k,\alpha^k)
 =
 w^k(E_\nu^k,\alpha^k)
 \frac{\Phi_{\alpha^k}(E_\nu^k) T}{\Phi_{\alpha^k,\rm{ref}}(E_\nu^k) T_{\rm ref}} \;.
\end{equation}
Our analysis uses events binned in $E_{\rm dep}$ and $\cos\theta_{z,\mathrm{rec}}$; we introduce our choice of binning in Sec.~\ref{sec:bounds_diffuse-stat_anal}.  In each bin, the mean expected number of events with topology t is the sum of the weights of MC events with topology t.

As in the 7.5-year HESE analysis by the IceCube Collaboration~\cite{IceCube:2020wum}, we include the irreducible background of atmospheric neutrinos and muons (Sec.~\ref{sec:bounds_diffuse-stat_anal}), and detector systematic uncertainties---the efficiency of digital optical modules, the head-on efficiency, and the lateral efficiency.  We keep the latter fixed to their nominal expectations from \Refe~\cite{IceCube:2020wum}.  As shown in \Refe~\cite{Fiorillo:2022rft}, doing this has a negligible impact on the reconstructed spectral properties of the flux, and allows us to considerably reduce the computational time needed for our analysis.


\subsection{Statistical analysis}
\label{sec:bounds_diffuse-stat_anal}

Our statistical analysis follows closely the one we used for NGC 1068 in Sec.~\ref{sec:bounds_ngc1068-stat}.  As discussed above (Sec.~\ref{sec:bounds_diffuse-flux}), we consider two scenarios of neutrino source populations, 1P and 2P.  In the 1P scenario, the diffuse neutrino flux from the single source population is determined by the set of astrophysical parameters (Sec.~\ref{sec:bounds_diffuse-flux}) $\boldsymbol{\theta}_{\rm{ast}}^{1{\rm P}}=\left\{\Phi_0, \gamma, f_{e, {\rm S}}\right\}$ (since we set the $\nu_\tau$ fraction at production to $f_{\tau, {\rm S}} = 0$, the $\nu_\mu$ fraction is $f_{\mu, {\rm S}} \equiv 1 - f_{e, {\rm S}}$), the mixing parameters $\boldsymbol{\theta}_{\rm{mix}}=\left\{\theta_{12},\theta_{23},\theta_{13},\delta_{\rm CP}\right\}$ and the decay parameters $\boldsymbol{\theta}_{\rm{dec}}=\left\{m_2/\tau_2, m_3/\tau_3\right\}$. For the 2P model, the set of astrophysical parameters is duplicated to account for two populations, \ie, $\boldsymbol{\theta}_{\rm{ast}}^{2{\rm P}}=\left\{\Phi_{0,1},\gamma_1,f_{e,{\rm S},1},\Phi_{0,2},\gamma_2,f_{e,{\rm S}, 2}\right\}$. 

For each choice of the above parameters, we use the MC sample (Sec.~\ref{sec:bounds_diffuse-hese_mc}) to generate the associated event sample, binned in reconstructed energy and direction. Similarly to \Refes~\cite{IceCube:2020wum,Fiorillo:2022rft}, we use $N_{E_{\rm dep}} = 21$ bins in energy, evenly spaced in $\log_{10}(E_{\rm dep}/{\rm GeV})$, between 60~TeV and 10~PeV, and $N_{c \theta_z^{\rm rec}} = 10$ bins in direction, evenly spaced  in $\cos \theta_z^{\rm rec}$, between -1 and 1. 
We denote by $\mu_{ij,\rm t}^{\nu,\rm{ast}}(\boldsymbol{\theta}_{\rm{ast}},\boldsymbol{\theta}_{\rm{mix}},\boldsymbol{\theta}_{\rm{dec}},T)$ the mean expected number of events with topology t induced by the astrophysical neutrino flux in the $i$-th energy bin and $j$-th angular bin, computed for the choice of parameters $\boldsymbol{\theta}_{\rm{ast}}$ (either $\boldsymbol{\theta}_{\rm{ast}}^{\rm 1P}$ or $\boldsymbol{\theta}_{\rm{ast}}^{\rm 2P}$), $\boldsymbol{\theta}_{\rm{mix}}$, and $\boldsymbol{\theta}_{\rm{dec}}$, and for an exposure time $T$.

To account for the contamination due to the irreducible atmospheric background, first, we extract from the IceCube HESE MC sample~\cite{IC75yrHESEPublicDataRelease} the baseline number of conventional atmospheric neutrinos, $N^{\nu, {\rm c}}_{ij, {\rm t}}(T)$, prompt atmospheric neutrinos, $N^{\nu, {\rm pr}}_{ij, {\rm t}}(T)$, and atmospheric muons, $N^{\mu}_{ij, {\rm t}}(T)$.  We keep the shape of the background energy and direction event distributions fixed, but allow their normalization constants, $\mathcal{N}^{\nu, {\rm c}}$, $\mathcal{N}^{\nu, {\rm pr}}$, and $\mathcal{N}^{\mu}$, to float independently of each other.  Thus, the number of background events of topology t is
\begin{eqnarray}
 \mu_{ij, {\rm t}}^{\rm atm}
 (\boldsymbol\eta,
 T)
 &=&
 \mathcal{N}^{\nu, {\rm c}}
 N^{\nu, {\rm c}}_{ij, {\rm t}}(T)
 +
 \mathcal{N}^{\nu, {\rm pr}} 
 N^{\nu, {\rm pr}}_{ij, {\rm t}}(T)
 \nonumber \\
 &&
 +~
 \mathcal{N}^{\mu} 
 N^{\mu}_{ij, {\rm t}}(T) \;,
\end{eqnarray}
where $\boldsymbol\eta \equiv (\mathcal{N}^{\nu, {\rm c}}, \mathcal{N}^{\nu, {\rm pr}}, \mathcal{N}^{\mu})$.

Altogether, the mean number of events of topology t in each bin, of astrophysical and atmospheric origin, is
\begin{eqnarray}
 \label{equ:mean_rate_total}
 \mu_{ij, {\rm t}}
 (\boldsymbol\theta_{\rm ast},
 \boldsymbol{\theta}_{\rm{mix}},
 \boldsymbol\theta_{\rm dec}, 
 \boldsymbol\eta,
 T)
 &=&
 \mu_{ij, {\rm t}}^{\nu, {\rm ast}}(\boldsymbol\theta_{\rm ast},
 \boldsymbol{\theta}_{\rm{mix}},
 \boldsymbol\theta_{\rm dec},
 T)
 \nonumber \\
 && 
 +~
 \mu_{ij, {\rm t}}^{\rm atm}(\boldsymbol\eta, T) \;.
\end{eqnarray}

Like for NGC 1068 before, we perform two analyses: first, using real HESE data and, second, using projections.  We describe them below.

\smallskip

\textbf{\textit{Using present-day data.---}}We compare the expected number of HESE events of each topology, $\mu_{ij, {\rm t}}$, against the number of events in the IceCube public 7.5-year sample, in each energy and direction bin, $n_{ij,\mathrm{t}}$.  To compare them, in analogy to \equ{likelihood_ngc1068}, we use a binned Poisson likelihood that spans all bins and topologies, \ie,
\begin{eqnarray}
 \label{equ:likelihood}
 &&
 \ln
 \mathcal{L}
 (\boldsymbol\theta_{\rm ast},
 \boldsymbol{\theta}_{\rm{mix}},
 \boldsymbol\theta_{\rm dec}, 
 \boldsymbol\eta,
 T)
 \\
 &&
 \qquad
 =
 \sum_{i=1}^{N_{E_{\rm dep}}}
 \sum_{j=1}^{N_{c\theta_z^{\rm rec}}}
 \sum_{\rm t}^{\{ {\rm c}, {\rm tr}, {\rm dc} \}}
 \ln
 \mathcal{L}_{ij,{\rm t}}
 (\boldsymbol\theta_{\rm ast},
 \boldsymbol{\theta}_{\rm{mix}},
 \boldsymbol\theta_{\rm dec}, 
 \boldsymbol\eta,
 T) 
 \nonumber
 \\
 &&
 \qquad \quad
 +~
 \frac{\chi^2(\boldsymbol{\theta}_{\rm{mix}})}{2}
 +
 \frac{\chi^2(\boldsymbol\eta)}{2}
 \nonumber
 \;,
\end{eqnarray}
where the likelihood in each bin, for topology t, is
\begin{equation}
 \mathcal{L}_{ij,{\rm t}}
 (\boldsymbol\theta_{\rm ast},
 \boldsymbol{\theta}_{\rm{mix}},
 \boldsymbol\theta_{\rm dec}, 
 \boldsymbol\eta,
 T)
 =
 \frac{\mu_{ij,\rm t}^{n_{ij,\rm t}}}{n_{ij,\rm t}!}
 e^{-\mu_{ij,\rm t}} \;, 
\end{equation}
and, on the right-hand side, $\mu_{ij,\rm t} \equiv \mu_{ij,\rm t}(\boldsymbol\theta_{\rm ast}, \boldsymbol{\theta}_{\rm{mix}},  \boldsymbol\theta_{\rm dec}, \allowbreak \boldsymbol\eta, T)$ is computed using \equ{mean_rate_total}.  In \equ{likelihood}, $\chi^2(\boldsymbol{\theta}_{\rm{mix}})/2$ and $\chi^2(\boldsymbol{\eta})/2$ are, respectively, pull terms that represent our prior knowledge of the mixing and atmospheric-background parameters.  We do not include an analogous term for the astrophysical parameters to avoid introducing bias in our results on the decay parameters.  For the mixing parameters, we use as priors their distributions from the recent global fit to oscillation data, NuFIT~5.2~\cite{Esteban:2020cvm,NuFit5.2}, assuming normal neutrino mass ordering.  For the atmospheric parameters, we use the same priors as the IceCube 7.5-year HESE analysis~\cite{IceCube:2020wum}, which are extracted from \Refe~\cite{Fedynitch:2012fs}.  

In analogy to \equ{likelihood_ngc1068_profiled}, we profile the likelihood function over all the nuisance parameters, \ie,
\begin{equation}
 \tilde{\mathcal{L}}
 (\boldsymbol{\theta}_{\rm dec},
 T)
 =
 \mathrm{max}_{\boldsymbol{\theta}_{\rm ast},\boldsymbol{\theta}_{\rm mix},\boldsymbol{\eta}}
 \mathcal{L}
 (\boldsymbol\theta_{\rm ast},
 \boldsymbol{\theta}_{\rm{mix}},
 \boldsymbol\theta_{\rm dec}, 
 \boldsymbol\eta,
 T) \;.
\end{equation}
Like before, the best-fit value of the decay parameters, $\hat{\boldsymbol{\theta}}_{\rm dec}=\left\{0,0\right\}$, corresponds to no decay of $\nu_2$ and $\nu_3$.  We compute lower bounds on the neutrino lifetimes $\tau_2/m_2$ and $\tau_3/m_3$, in analogy to \equ{test_statistic_ngc1068}, via the test statistic
\begin{equation}
 \Lambda(\boldsymbol{\theta}_\mathrm{dec}, T)
 =
 2
 \left[
 \log\tilde{\mathcal{L}}(\hat{\boldsymbol{\theta}}_{\rm dec}, T)
 -
 \log\tilde{\mathcal{L}}({\boldsymbol{\theta}_{\rm dec}}, T)
 \right] \;.
\end{equation}
And, again, we report two-dimensional confidence intervals in $\tau_2/m_2$ and $\tau_3/m_3$ based on Wilks' theorem. 

\smallskip

\textbf{\textit{Using projected data.---}}As discussed in Sec.~\ref{sec:nu_prop_with_decay-diffuse_vs_point}, the sensitivity to neutrino decay from the diffuse flux relies predominantly on the flavor information.  The larger event samples that will be made possible by upcoming detectors will provide a quantitative and qualitative change in our measurement of the flavor composition, pinpointing it to unprecedented precision~\cite{Song:2020nfh} and potentially shedding light on energy-dependent flavor effects~\cite{Liu:2023flr}.  We explore how much this will improve the sensitivity to neutrino decay. 

To answer this question, we generate samples of HESE events detected collectively by upcoming neutrino telescopes, \`a la PLE$\nu$M~\cite{Schumacher:2021hhm}.  Starting in 2025, we consider, in addition to IceCube, the exposure of Baikal-GVD~\cite{Baikal-GVD:2020xgh, Baikal-GVD:2020irv} (assumed effective volume of 1.5 times IceCube) and KM3NeT~\cite{KM3Net:2016zxf, Margiotta:2022kid} (2.8 times IceCube). After 2030, we replace IceCube for IceCube-Gen2~\cite{IceCube-Gen2:2020qha} (8 times IceCube) and consider the additional exposure of P-ONE~\cite{P-ONE:2020ljt} (3.2 times IceCube), TAMBO~\cite{Romero-Wolf:2020pzh, TAMBO:2023plw} (0.5 times IceCube), and TRIDENT~\cite{Ye:2022vbk} (7.5 times IceCube).  The timelines and effective volumes for each of the detectors are summarized in Fig.~1 of \Refe~\cite{Fiorillo:2022rft}.  They are to be taken as tentative, since they are subject to change.  

We assume that each detector has a different size than IceCube, but the same HESE-detection capabilities as IceCube.  This is admittedly a necessary simplification due to the absence of details on future detectors.  (This may be improved in the future by using dedicated detector simulation tools, \eg, \Refe~\cite{Lazar:2023rol}.)  The one exception is TAMBO; since it is meant as a $\nu_\tau$ detector, for it we only consider events generated by the $\nu_\tau$ flux.    We compute the projected combined HESE event rates for the year 2035, which correspond to an equivalent IceCube exposure of $T_{\rm proj} = 159$~yr.

To generate projected HESE samples, we assume the 1P source population model, computed using the star-formation rate for the source redshift evolution, and fix the model parameters to their present-day best-fit values: $\hat{\boldsymbol{\theta}}_{\rm ast}$ and $\hat{\boldsymbol{\eta}}$, from the analysis of the 7.5-year HESE sample by the IceCube Collaboration~\cite{IceCube:2020wum}; $\hat{\boldsymbol{\theta}}_{\rm mix}$, from NuFIT~5.2~\cite{Esteban:2020cvm, NuFit5.2}; and $\hat{\boldsymbol{\theta}}_{\rm decay} = \{ 0, 0 \}$, corresponding to no neutrino decay.  We use \equ{mean_rate_total} to generate a mock Asimov event sample~\cite{Cowan:2010js}, where the observed values are identical to the expected values, \ie, we set $n_{ij,\mathrm{t}}=\mu_{ij,\mathrm{t}}(\hat{\boldsymbol{\theta}}_{\rm ast},\hat{\boldsymbol{\theta}}_{\rm mix},\hat{\boldsymbol{\theta}}_{\rm dec},\hat{\boldsymbol{\eta}}, T_{\rm proj})$.  Like before, we use the likelihood function, \equ{likelihood}, to compare this against test event rates, $\mu_{ij,\mathrm{t}}(\boldsymbol{\theta}_{\rm ast}, \boldsymbol{\theta}_{\rm mix}, \boldsymbol{\theta}_{\rm dec}, \boldsymbol{\eta},  T_{\rm proj})$.  We use the same priors on the mixing and atmospheric parameters as for our results based on present-day data. 

By maintaining the present-day priors on the mixing parameters, we neglect the expected improvement in their precision made possible by upcoming oscillation experiments, which should be significant; see, \eg, Fig.~1 in \Refe~\cite{Song:2020nfh}.  By 2035, the uncertainty on the mixing parameters should be small enough to render its impact on the expected flavor composition at Earth negligible (see Fig.~2 in \Refe~\cite{Song:2020nfh}), allowing for cleaner tests of neutrino decay.  Yet, the precise ranges of the neutrino lifetimes that could be excluded will depend on the \textit{real} values of the mixing parameters, which are presently unknown, but are contained within the range presently allowed by experiments.  Thus, maintaining the present-day priors on the mixing parameters makes our projected bounds on neutrino decay conservative.


\subsection{Results}

Figure~\ref{fig:bounds_diffuse} shows our resulting two-dimensional joint bounds on the lifetimes of $\nu_2$ and $\nu_3$, present and projected, for the scenarios listed in Table~\ref{tab:diffuse}.  Similarly to our results from NGC 1068, this table, and \figu{1d_bounds_summary}, show also the one-dimensional bounds on the individual lifetimes that, however, discard important correlations between $\tau_2/m2$ and $\tau_3/m_3$, as we explain below. 

Today, because of the limited number of HESE events, only rather large deviations in the flavor composition relative to its standard expectation can be disfavored.  This means that only the case where both $\nu_2$ and $\nu_3$ decay can be disfavored (at 95\%~C.L.), since it leads to the largest decay-induced deviations (\figu{flavor_triangle}).  In \figu{bounds_diffuse}, this is reflected in the correlation between the disfavored regions of $\nu_2$ and $\nu_3$ lifetimes.  The bounds come from the fact that the observed flavor composition at Earth cannot be reproduced by \textit{any} choice of flavor composition of the form $(f_{e,{\rm S}}, 1-f_{e, {\rm S}}, 0)$ at the source if both $\nu_2$ and $\nu_3$ decay sufficiently rapidly.

The strongest bounds are obtained assuming SFR evolution of the sources, since it places sources farther away; roughly, they exclude lifetimes of  $\tau_2/m_2 \lesssim 500$~s~eV$^{-1}$ and $\tau_3/m_3 \lesssim 100$~s~eV$^{-1}$.  Assuming instead BL Lac evolution weakens the bounds by about one order of magnitude, because more sources are closer to Earth.  For both choices of source evolution, the bounds are impervious to switching between using one (1P) or two (2P) source populations, since the inability to explain the flavor composition at Earth under decay holds regardless of the number of source populations.

In our 2035 projections, the disfavored regions expand in two  ways.  First, the projected bounds exclude longer $\nu_3$ lifetimes.  This is because the projected increase in the event rates allows us to look for decay-induced flavor transitions even if they happen within 1--10~PeV, where the number of detected neutrinos is comparatively lower; this is in agreement with \Refe~\cite{Liu:2023flr}.  Second, the projected bounds become essentially sensitive only to $\tau_3/m_3$; even if $\nu_2$ were stable, the decay of $\nu_3$ would still be excluded. This ties in with our discussion of \figu{flavor_triangle} in Sec.~\ref{sec:nu_prop_with_decay-flavor_comp}, where we showed that an improved measurement of the flavor composition would allow us to eliminate entirely the possibility that $\nu_3$ decays. 

Beyond this, the projected bounds, like the present-day ones, depend on the assumed evolution of the sources, with bounds computed under BL Lac evolution 3--10 times weaker than bounds computed under SFR evolution, depending on the value of the $\nu_2$ lifetime.  On the other hand, while the differences between the 1P and 2P scenarios are more visible than in the present day, they remain relatively mild. These differences only appear in the region where the transition between no decay and full decay in the spectrum (\figu{flux_diffuse}) is in the middle of the IceCube energy range. In this case, the effect of decay is visible not just in the flavor composition, but also in the shape of the spectrum, which depends on the choice between 1P and 2P.  Because the 2P scenario has a larger parameter space than the 1P scenario, the bounds derived under it are slightly weaker. 


\subsection{Future improvements}
\label{sec:bounds_diffuse-future}

Our projections are based on the measurement of the diffuse flux using exclusively HESE events.  Combining measurements of the diffuse flux via HESE events and  through-going tracks~\cite{IceCube:2021uhz, IceCube:2023hou} would boost the precision with which the diffuse $\nu_\mu$ content is measured~\cite{IceCube:2015gsk, Song:2020nfh, Naab:2023xcz, Liu:2023flr}.  In our analysis, we only considered HESE events because there are no publicly available tools provided by the IceCube Collaboration that would allow us to treat through-going tracks with the same level of detail with which we treat HESE events.  However, \Refes~\cite{Shoemaker:2015qul, Song:2020nfh} showed that there is potential for improvement in the sensitivity to neutrino decay from such combined analyses.  A full analysis combining HESE and tracks events remains to be performed, but lies beyond the scope of this work.

Further, we have assumed that future neutrino telescopes will have detection capabilities equal to the present-day capabilities of IceCube, the improvement in our lifetime bounds due only to the increase in their combined detection rate. In reality, the improvement may be hastened by ongoing progress in event reconstruction aided by machine learning~\cite{IceCube:2021umt, IceCube:2023ame, IceCube:2022njh, Sogaard:2022qgg, Bukhari:2023ezc} and by new techniques in flavor identification, like dedicated templates~\cite{IceCube:2020fpi, IceCube:2023fgt} and muon and neutron echoes~\cite{Li:2016kra, Steuer:2017tca, Farrag:2023jut}.  


\section{Summary and outlook}
\label{sec:discussion}

In the Standard Model, neutrinos are effectively stable.  Therefore, discovering their decay would constitute evidence of new physics.  We have searched for signs of neutrino decay in TeV--PeV astrophysical neutrinos, whose cosmological-scale baselines, in the $L = $~Mpc--Gpc range, make them sensitive to decay even if their lifetimes are of millions or billions of years.

We have explored a generic benchmark scenario of \textit {invisible} neutrino decay of the neutrino mass eigenstates $\nu_2$ and $\nu_3$ decay into undetected particles, while $\nu_1$ is stable.  Their decay distorts the energy distribution of the flux high-energy astrophysical neutrinos that reach Earth and its flavor composition (Sec.~\ref{sec:nu_prop_with_decay}).  The size of the distortion depends on the factors $e^{-\frac{m_2}{\tau_2} \frac{L}{E_\nu}}$ and $e^{-\frac{m_3}{\tau_3}\frac{L}{E_\nu}}$, where $E_\nu$ is the neutrino energy, and $m_i$ and $\tau_i$ are the mass and lifetime of $\nu_i$ ($i = 1, 2$).  We constrain the ``lifetimes'' of $\nu_2$ and $\nu_3$, $\tau_2/m_2$ and $\tau_3/m_3$, by contrasting our predictions of the neutrino flux against experimental data.  

The main goal of our work is to provide a fresh perspective on searching for neutrino decay using high-energy astrophysical neutrinos.  We improve upon previous studies in two key, but often overlooked or understudied aspects.  

First, we explore broadly the impact of the large astrophysical unknowns that plague the modeling of the high-energy neutrino flux and that cloud signs of decay.  This includes the shape of the neutrino energy distribution, the flavor composition with which neutrinos are emitted, the redshift distribution of astrophysical neutrino sources, the number of source populations, and whether or not we have strong priors on the size of neutrino flux. 

Second, we include the effects of realistic neutrino detection capabilities by means of detailed detector simulations, which aggravates the issue by weakening signs of decay due to the limited precision with which the neutrino energy, direction, and flavor are inferred.  The impact of the above ranges from appreciable to critical; ignoring it would yield unrealistically high sensitivity.  

We have applied the above perspectives to constraining neutrino lifetimes using, for the first time, the flux of high-energy neutrinos from the recently discovered steady-state source candidate, the active galaxy NGC 1068, and also using the diffuse flux of high-energy astrophysical neutrinos, \ie, the aggregated flux of all unresolved neutrino sources.  Our constraints are based on present-day neutrino data collected by the IceCube neutrino telescope and on projected data, for the year 2035, collected jointly by IceCube and upcoming telescopes. 

\textit{\textbf{Using neutrinos from NGC 1068 (Sec.~\ref{sec:bounds_ngc1068}), we find that, today, it is not possible to constrain neutrino decay.}}  At first, this result is somewhat surprising, since in this case we know the distance to the single source from which neutrinos originate---about 14 Mpc---and so, naively, we would expect to constrain lifetimes of size $\tau_i/m_i \simeq L/E_\nu$.  The main reason why in practice this is not the case is that, given the large uncertainties in models of neutrino emission from NGC 1068, the precision with which IceCube currently measures the neutrino flux is insufficient to spot the factor-of-two-thirds change in the flux normalization, at most, that decay may introduce (Sec.~\ref{sec:nu_prop_with_decay-energy_spectrum}).  However, our projections (\figu{bounds_ngc1068}) show that future telescopes could disfavor lifetimes of 100--5000~s~eV$^{-1}$ (95\%~C.L.).

\textit{\textbf{Using the diffuse neutrino flux (Sec.~\ref{sec:bounds_diffuse}), we place competitive lower bounds on neutrino lifetimes already today}}, using the public IceCube 7.5-year HESE sample (\figu{bounds_diffuse}).  Lifetimes roughly in the range of 20--450~s~eV$^{-1}$ are disfavored (at 95\%~C.L.), yet the lifetimes of $\nu_2$ and $\nu_3$ cannot be constrained individually, only jointly (unless they are assumed to be equal, see below). These are arguably the most robust limits garnered from high-energy astrophysical neutrinos to date.  Our projections show an order-of-magnitude improvement in the bounds by 2035.

We recommend using the preferred form of our constraints, jointly on $\tau_2/m_2$ and $\tau_3/m_3$ (Figs.~\ref{fig:bounds_ngc1068} and \ref{fig:bounds_diffuse}), which can be found in \Refe~\cite{github_repo}.  Although we also report one-dimensional constraints on the lifetimes (Tables~\ref{tab:ngc_1068} and \ref{tab:diffuse}, and Figs.~\ref{fig:bounds_compare} and \ref{fig:1d_bounds_summary}), we caution against taking them at face value, since the regions of disfavored $\nu_2$ and $\nu_3$ lifetimes are correlated, especially in the case of NGC 1068.  Likewise, we caution against using constraints obtained assuming that the $\nu_2$ and $\nu_3$ lifetimes are equal---a common assumption in the literature---since doing so may convey sensitivity to decay in cases where there is no sensitivity to the individual lifetimes; see, \eg, the present-day bounds from the diffuse flux in Table~\ref{tab:diffuse}.

The bounds on neutrino lifetime that we report outclass bounds inferred from solar, atmospheric, accelerator, and reactor neutrinos; see \figu{bounds_compare}.  However, future bounds inferred from detecting the neutrinos from the next Galactic supernova will in turn outclass ours by several orders of magnitude~\cite{Martinez-Mirave:2024hfd}, on account of their Lorentz boost being weaker due to their lower energies.  

While we wait for the next Galactic supernova to occur, however, high-energy astrophysical neutrinos continue to be powerful probes of neutrino decay.


\section*{Acknowledgements}

We thank Pablo Martìnez Miravè and Kohta Murase for illuminating discussion and, especially, Lisa Schumacher for help with the PLE$\nu$M code.  DF, MB, and VBV are supported by the {\sc Villum Fonden} under project no.~29388.  This work used the Tycho supercomputer hosted at the SCIENCE High Performance Computing Center at the University of Copenhagen. 


\bibliography{refs}


\newpage
\appendix


\section{One-dimensional lifetime bounds}
\label{app:1d_limits}

\renewcommand{\theequation}{A\arabic{equation}}
\renewcommand{\thefigure}{A\arabic{figure}}
\renewcommand{\thetable}{A\arabic{table}}
\setcounter{figure}{0} 
\setcounter{table}{0} 

\begin{figure*}[!b]
 \includegraphics[width=\textwidth]{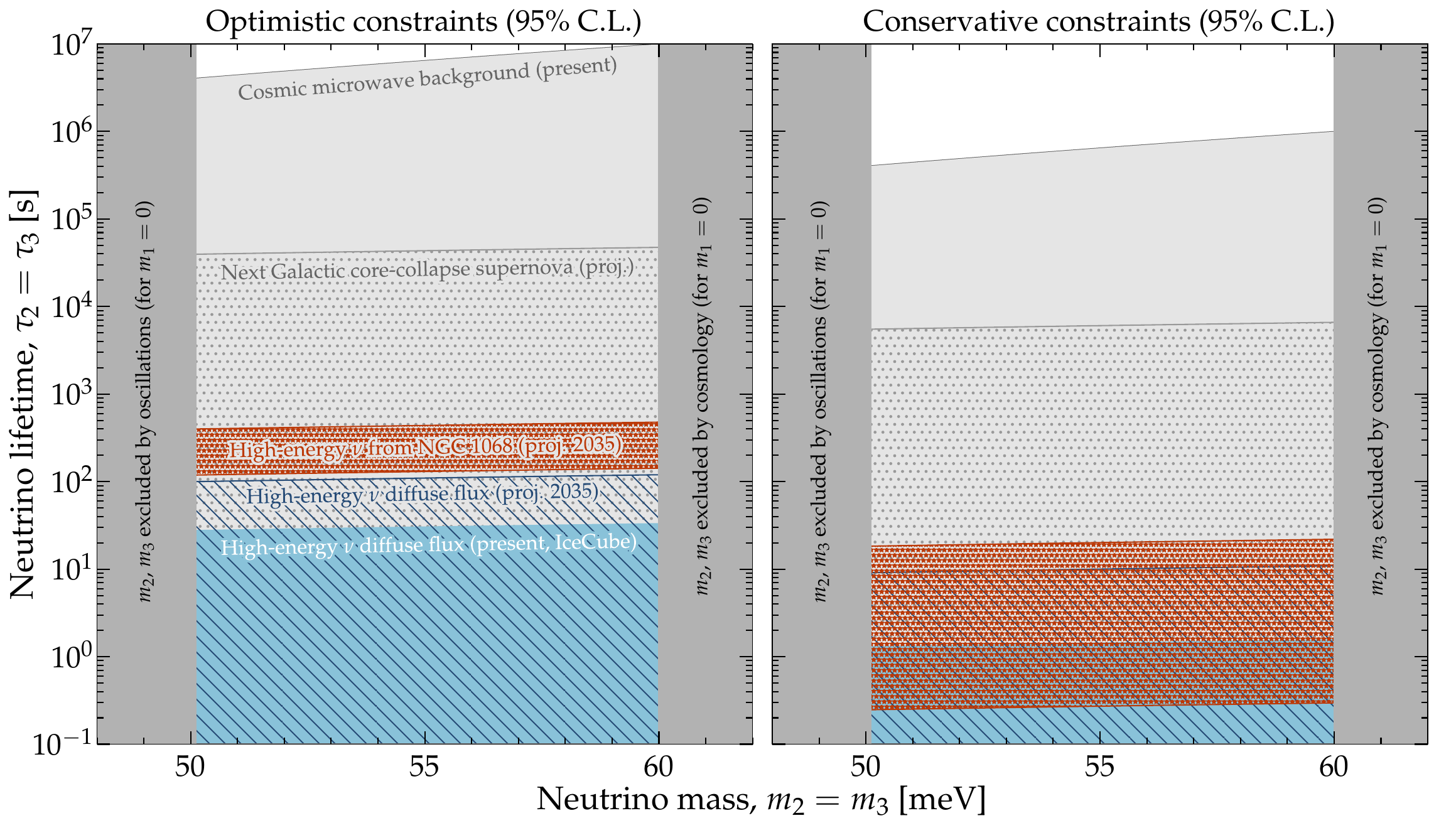}
 \caption{\textbf{\textit{Comparison between one-dimensional marginalized bounds on the neutrino lifetime inferred from high-energy astrophysical neutrinos and competitive bounds from the literature.}}  Shaded regions are disfavored at 95\%~C.L.  The bounds are obtained assuming the invisible decay of $\nu_2$ and $\nu_3$, with a common lifetime $\tau_2/m_2 = \tau_3/m_3$, while $\nu_1$ is stable.  \textit{Left:} bounds obtained under optimistic assumptions.   \textit{Right:} bounds obtained under conservative assumptions.  The bounds from the cosmic microwave background are from \Refe~\cite{Barenboim:2020vrr} (see also \Refe~\cite{Chen:2022idm}); the bounds from the next Galactic core-collapse supernova, from \Refe~\cite{Martinez-Mirave:2024hfd}.  The lower limit on $m_2$ and $m_3$ come from the measurement of the mass-squared differences, $\Delta m_{21}^2 \equiv m_2^2 - m_1^2$ and $\Delta m_{31}^2 \equiv m_3^2 - m_1^2$, in oscillation experiments; we use their best-fit values from NuFIT~5.2~\cite{Esteban:2020cvm, NuFit5.2}.  The upper limit on them comes from the upper limit on the sum of the neutrino masses, $m_1 + m_2 + m_3 \leq 0.12$, from cosmology~\cite{Planck:2018vyg}.  In both cases, we set $m_1 = 0$ to display the widest possible mass ranges, as in Fig.~1 of \Refe~\cite{Bustamante:2016ciw}.  See \figu{1d_bounds_summary} for an overview of all one-dimensional bounds, and Figs.~\ref{fig:bounds_ngc1068} and \ref{fig:bounds_diffuse} for the two-dimensional bounds on $\tau_2/m_2$ and $\tau_3/m_3$, no longer assuming that they are equal.  See Secs.~\ref{sec:bounds_ngc1068} and \ref{sec:bounds_diffuse} for details.}
 \label{fig:bounds_compare}
\end{figure*}

Figure~\ref{fig:bounds_compare} shows a comparison between our one-dimensional bounds~{\it vs.}~other relevant bounds from the literature.  See also Fig.~1 in \Refe~\cite{Bustamante:2016ciw}, Fig.~5 in \Refe~\cite{Arguelles:2022tki}, and Fig.~15 in \Refe~\cite{Martinez-Mirave:2024hfd}.  We show our bounds obtained under optimistic and conservative assumptions. 
\begin{description}
 \item[Optimistic bounds]
  The bound from the diffuse neutrino flux assumes a single source population (1P), fixed pion-beam flavor composition, and SFR source evolution, using present-day IceCube data (the 7.5-year IceCube HESE sample~\cite{IceCube:2020wum, IC75yrHESEPublicDataRelease}) and forecasts for the year 2035.  The bound from NGC 1068, viable only for the year 2035, assumes a power-law neutrino spectrum (PL), fixed pion-beam composition, and a tight prior on the normalization of the neutrino flux.
 \item[Conservative bounds]
  The bound from the diffuse neutrino flux assumes two source populations (2P), free flavor composition, and BL Lac source evolution.  The bound from NGC 1068 assumes a power-law neutrino spectrum with a high-energy cut-off (PC), fixed pion-beam composition, and no prior on the normalization of the neutrino flux. 
\end{description}

\begin{figure*}[t!]
 \centering
 \includegraphics[width=\textwidth]{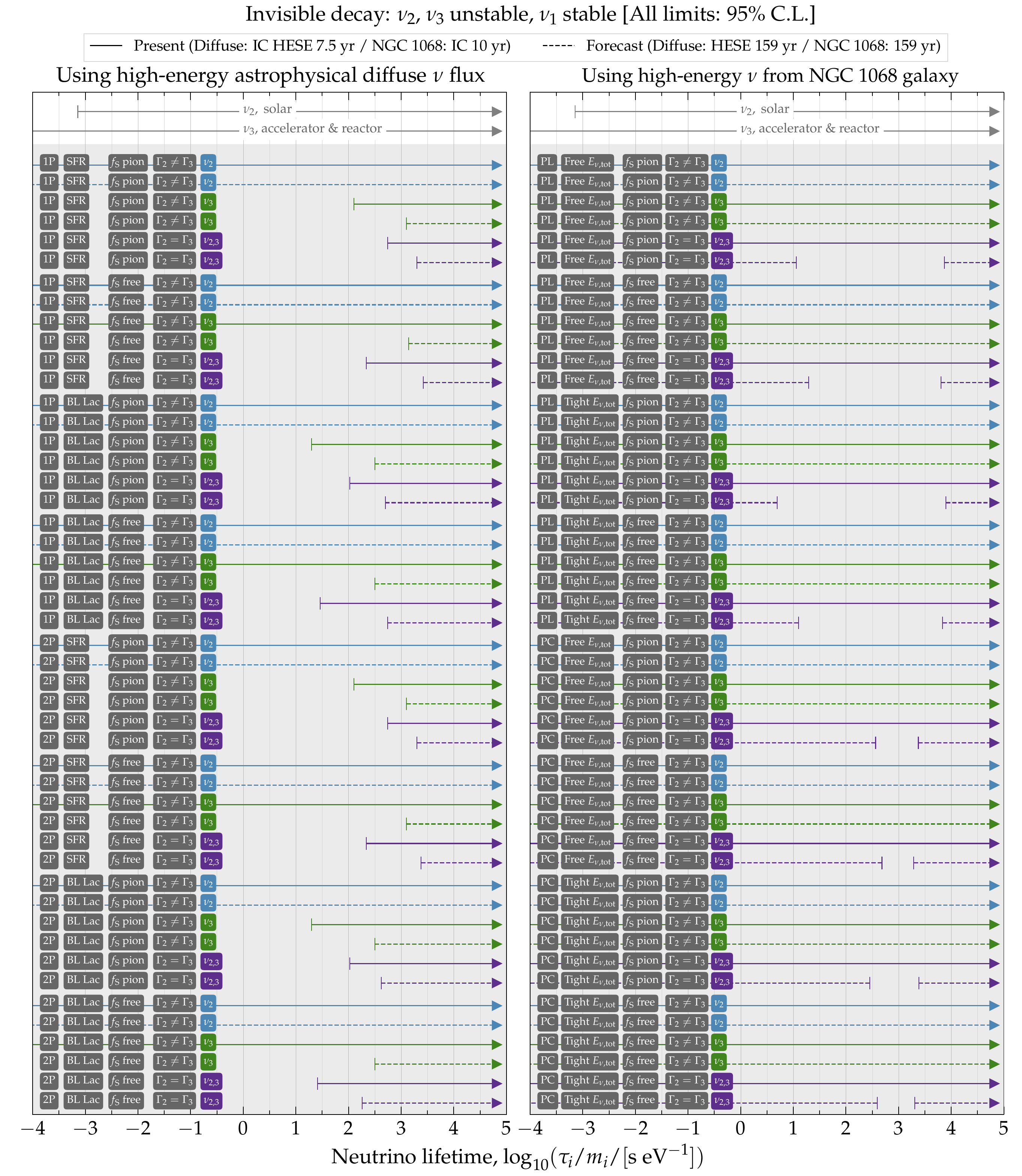}
 \caption{\textbf{\textit{Overview of one-dimensional bounds on the lifetimes of $\nu_2$ and $\nu_3$.}} We show present-day and year-2035 one-dimensional bounds separately on $\tau_{\nu_2}/m_{\nu_2}$ and $\tau_{\nu_3}/m_{\nu_3}$, profiled over all other model parameters, for all the different scenarios that we have explored. 
 We show bounds assuming free flavor composition at the sources (``$f_{\rm S}$ free'') or the composition from pion decay (``$f_{\rm S}$ pion''), letting $\nu_2$ and $\nu_3$ have different decay rates ($\Gamma_2 \neq \Gamma_3$) or a common one ($\Gamma_2 = \Gamma_3$).  
 {\it Left:} Bounds from the diffuse high-energy neutrino flux, assuming one (1P) or two (2P) neutrino source populations that evolve with redshift following the star-formation rate (SFR) or BL Lacs. 
 {\it Right:} Bounds from neutrinos from NGC 1068, assuming either a power-law neutrino spectrum (``PL'') or a power law with a high-energy exponential cut-off (``PC''), and either no constraint on the total emitted energy in neutrinos (``Free ``$E_{\nu, {\rm tot}}$'') or constraining it to be at most 50\% of the energy in x-rays (``Tight ``$E_{\nu, {\rm tot}}$'').
 }
 \label{fig:1d_bounds_summary}
\end{figure*}

Figure~\ref{fig:1d_bounds_summary} shows a comparison among all the one-dimensional bounds on the neutrino lifetimes, $\tau_2/m_2$ and $\tau_3/m_3$, obtained after profiling over all the other model parameters.  The numerical values are in Tables~\ref{tab:ngc_1068} and \ref{tab:diffuse}.

\onecolumngrid

\end{document}